\documentclass[a4paper,11pt]{article}
\usepackage{pos}
\usepackage{url}
\usepackage{subfigure}

\title{The past and future 20-years endeavor for discovering origins of ultra-high energy cosmic rays \\-- Rapporteur's summary of cosmic ray indirect --}
 \ShortTitle{Rapporteur's summary of CRI in ICRC2023}

\author{Toshihiro Fujii}

\affiliation[]{Graduate School of Science, Osaka Metropolitan University, Sumiyoshi, Osaka 558-8585, Japan}

\affiliation[]{Nambu Yoichiro Institute of Theoretical and Experimental Physics,\\
Osaka Metropolitan University, Sumiyoshi, Osaka 558-8585, Japan}

\emailAdd{toshi@omu.ac.jp}

\abstract{This article is the rapporteur's summary of the cosmic ray indirect sessions of the 38th International Cosmic Ray Conference in Nagoya, Japan. The rapporteur highlights cosmic ray indirect observatories around the world, and reviews a selection of the latest results regarding the cosmic ray energy spectrum, mass composition, anisotropy, hadronic interaction models, theory, geophysics, interdisciplinary research, and future projects.
\\\\\\ 
Published in T.~Fujii, PoS (ICRC2023) 031\,\,\,~\url{https://pos.sissa.it/444/031/}
}

\ConferenceLogo{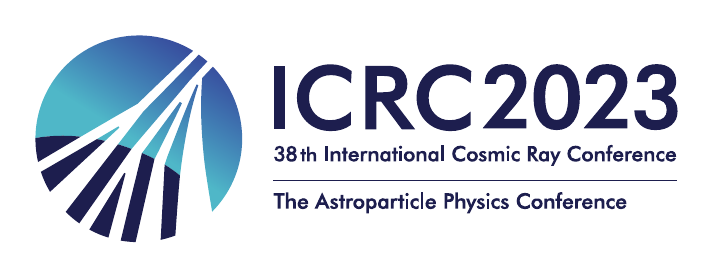}

\FullConference{%
38th International Cosmic Ray Conference (ICRC2023)\\
  26 July - 3 August, 2023\\
  Nagoya, Japan}

\begin{document}
\maketitle

\section{Receiving a baton as ``rapporteur in Japan''}
\begin{figure}[b]
    \centering
    \includegraphics[width=1.00\textwidth]{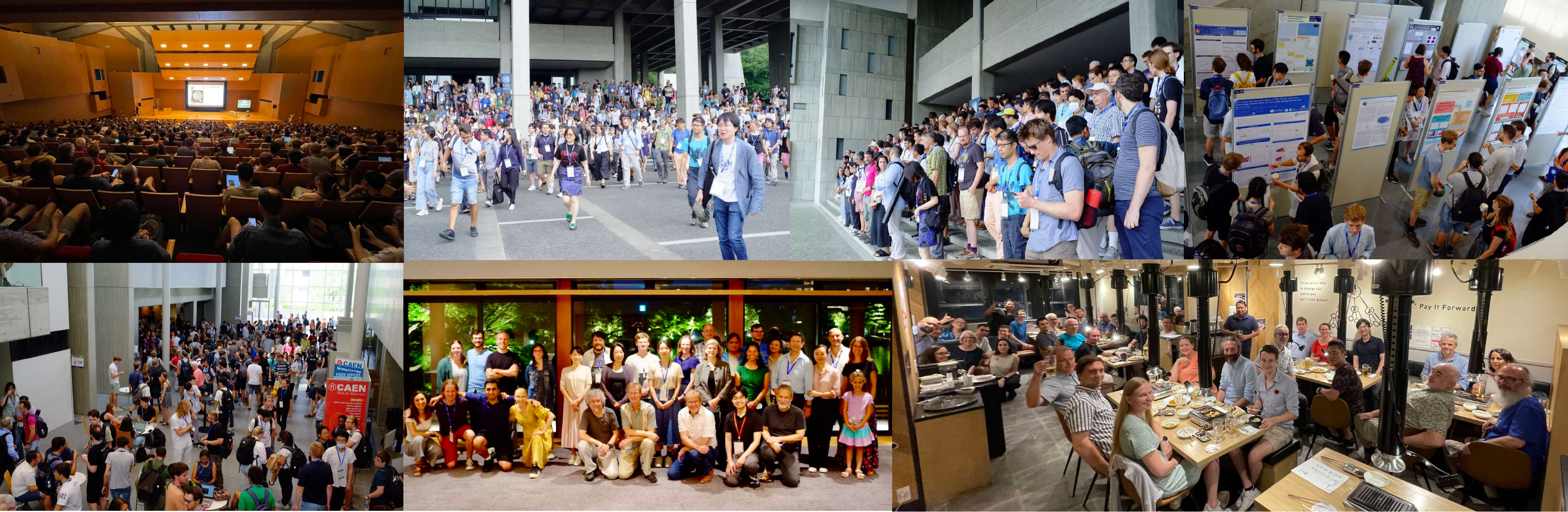}
    \caption{\textbf{Photos at ICRC2023.} These photos are from the opening session, photo session, poster session, coffee break, the Chicago dinner and Global Cosmic ray Observatory (GCOS) dinner. In-person events such as these, which were impossible during the pandemic, are important and allow for productive discussion and networking.} 
    \label{fig:photo}
\end{figure}

In the sweltering heat of Japan's hottest season, the 38th International Cosmic Ray Conference (ICRC2023) was held in Nagoya at Nagoya University. With more than one thousand enthusiastic on-site participants, ICRC2023 was the first in person ICRC held since the beginning of the COVID-19 pandemic. To enable maximum participation the conference was streamed online, making it the first hybrid (both onsite and online) ICRC. A glimpse of the conference activities is shown in Figure~\ref{fig:photo}.

In this proceedings, I summarize contributions from the ``cosmic ray indirect'' (CRI) sessions,
constituting 128 oral presentations, 211 posters, and 6 related plenary talks.
Although challenging (and exhausting), it was a great opportunity to discover not just the latest scientific results, but also the next-generation of cosmic ray scientists.
The detailed and intriguing discussions I had with students and younger members in the field left me feeling confident in the future of cosmic ray research.
I sincerely thank the contributors to the CRI sessions for your productive and fruitful discussions.
I would also like to express my deepest appreciation to the local organizing committee (LOC) of ICRC2023
for their hard work and allowing me to focus on my role as rapporteur.

\begin{figure}[t]
    \centering
    \subfigure[With Prof.~Angela Olinto]{\includegraphics[width=0.4\textwidth]{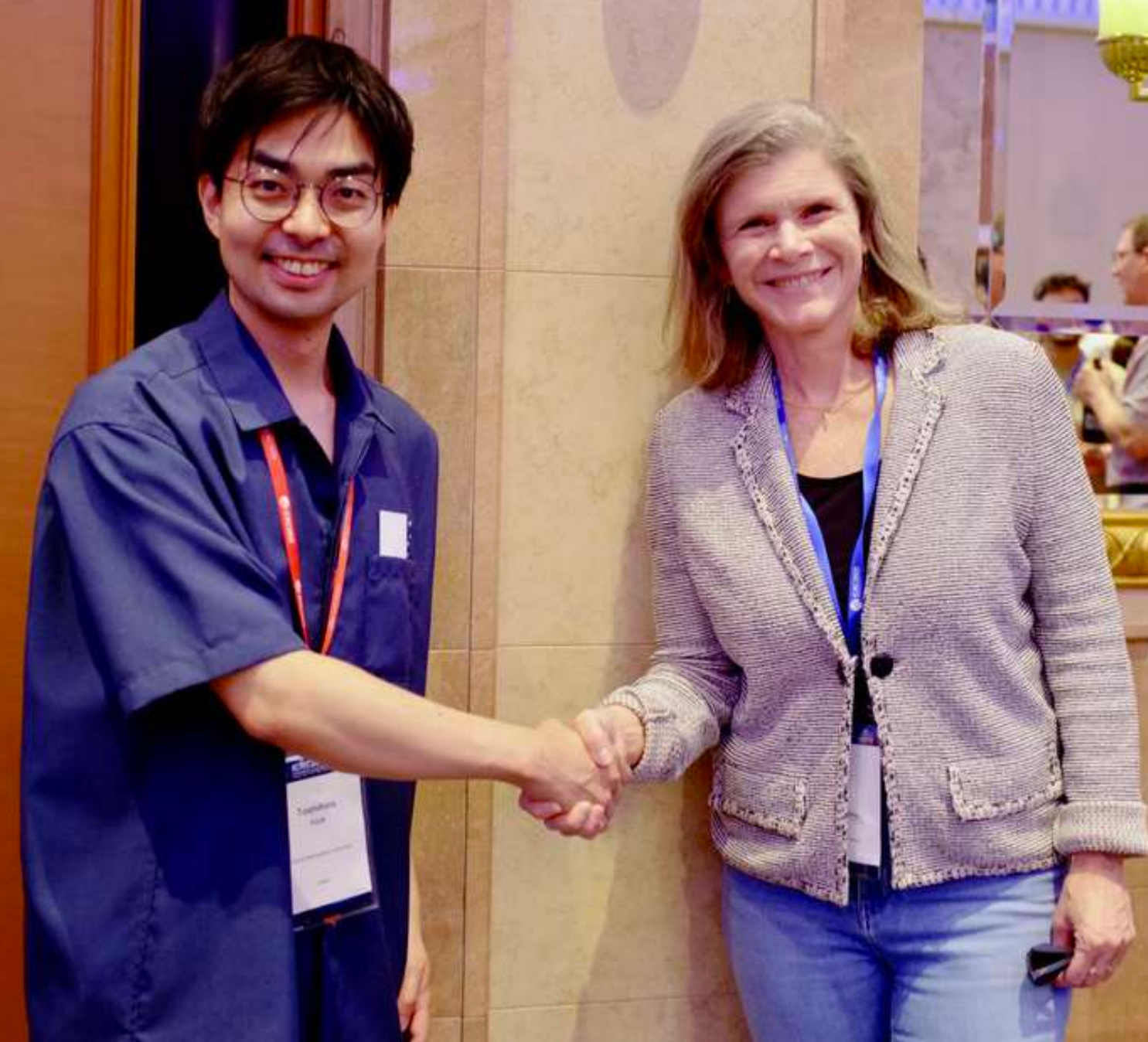}}
    \subfigure[Schematic view of UHECR astronomy]{\includegraphics[width=0.57\textwidth]{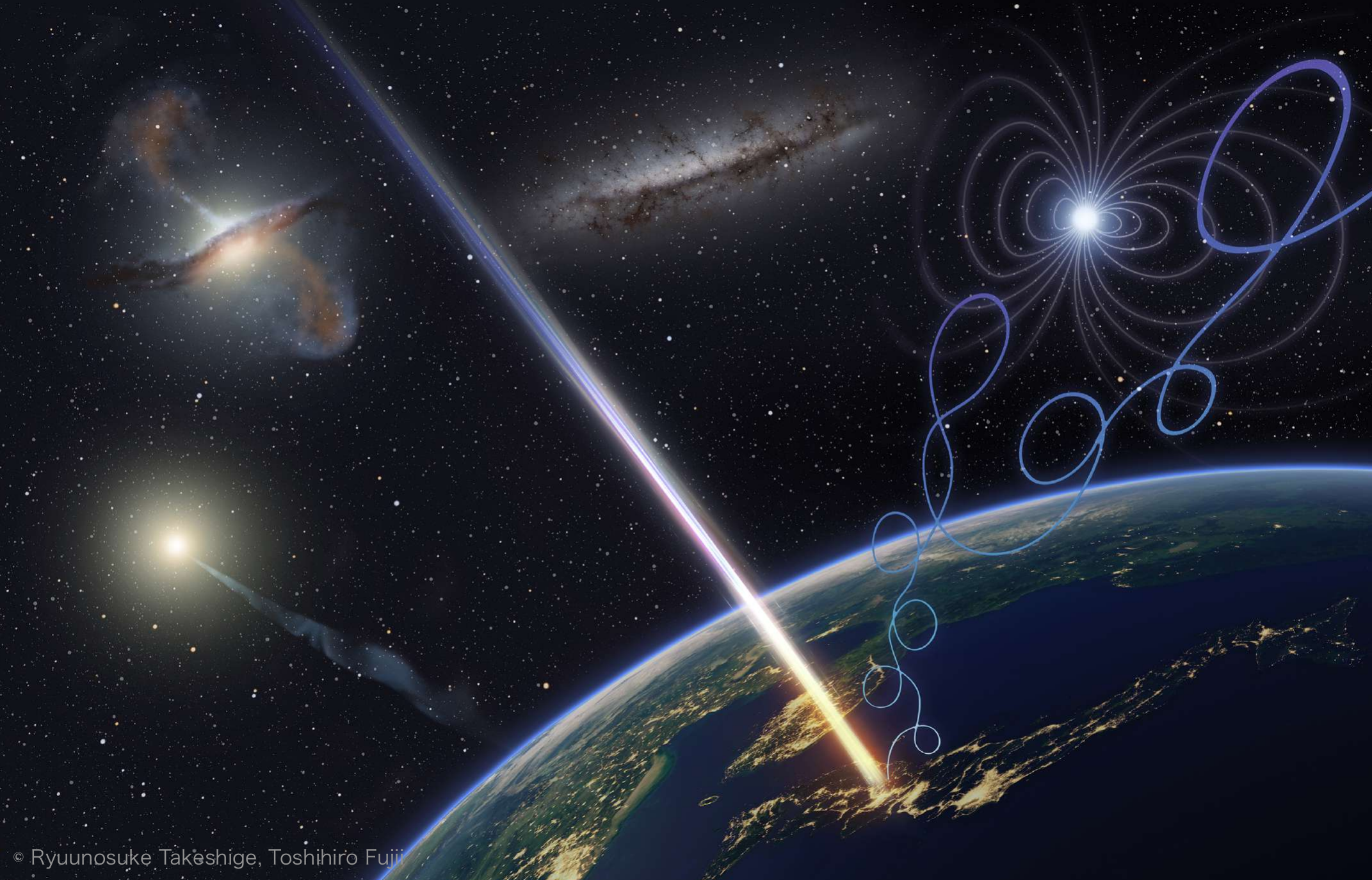}}
    \caption{\textbf{Receiving a baton as rapporteur in Japan and schematic view of UHECR astronomy.} (A) Memorial photo with Prof.~Angela Olinto, rapporteur of ICRC2003\@. (B) Conceptual image to indicate UHECR astronomy. The background image shows possible UHECR source candidates, such as active galactic nuclei, starburst galaxies and neutron stars.}
    \label{fig:baton}
\end{figure}

In April 2023, I received an email informing me that I had been nominated as the rapporteur of the ICRC2023 CRI session. 
I was flabbergasted to learn that I would have to summarize $>$340 contributions, in addition to assisting the LOC\@.
At the opening reception I met Prof.~Angela Olinto, who was the CRI rapporteur of ICRC2003 in Tsukuba, Japan.
She told me that ``it is a great honor'' and ``it is up to you how you do it''.
With her encouragement, I received a baton as ``rapporteur in Japan'' (Figure~\ref{fig:baton}A)\@.
At ICRC2023, there were nearly twice as many contributions as ICRC2003~\cite{Olinto:2004hc}, indicating a significant extension of the research field and increase in the number of active scientists.
%Also at a dinner\footnote{Conventionally the Chicagoan has a social dinner every ICRC, dubbed ``Chicago dinner''}, 

In this proceedings, I would like to describe a selection of results including my personal thoughts and future perspectives for ``passing the baton'' to the next generation of scientists. Furthermore, just as the Japanese people were told during the pandemic to consider the ``Three Cs'' (\textbf{C}losed spaces,
\textbf{C}rowded places and \textbf{C}lose-contact settings\footnote{\url{https://www.kantei.go.jp/jp/content/000061935.pdf}}), I would like to emphasize the importance of a different set of ``Three Cs'' in relation to cosmic ray research: \textbf{C}alibration, \textbf{C}ross check and \textbf{C}ollaboration.
The proceedings starts by briefly looking back at the pioneering work in cosmic ray observation performed in Nagoya, before introducing current CRI experiments across the world.
The latest energy spectrum, mass composition and anisotropy results, as well as hadronic interaction models, geophysics, interdisciplinary research, theory and future projects are discussed.
\begin{figure}[b]
    \centering
    \includegraphics[width=1.0\textwidth]{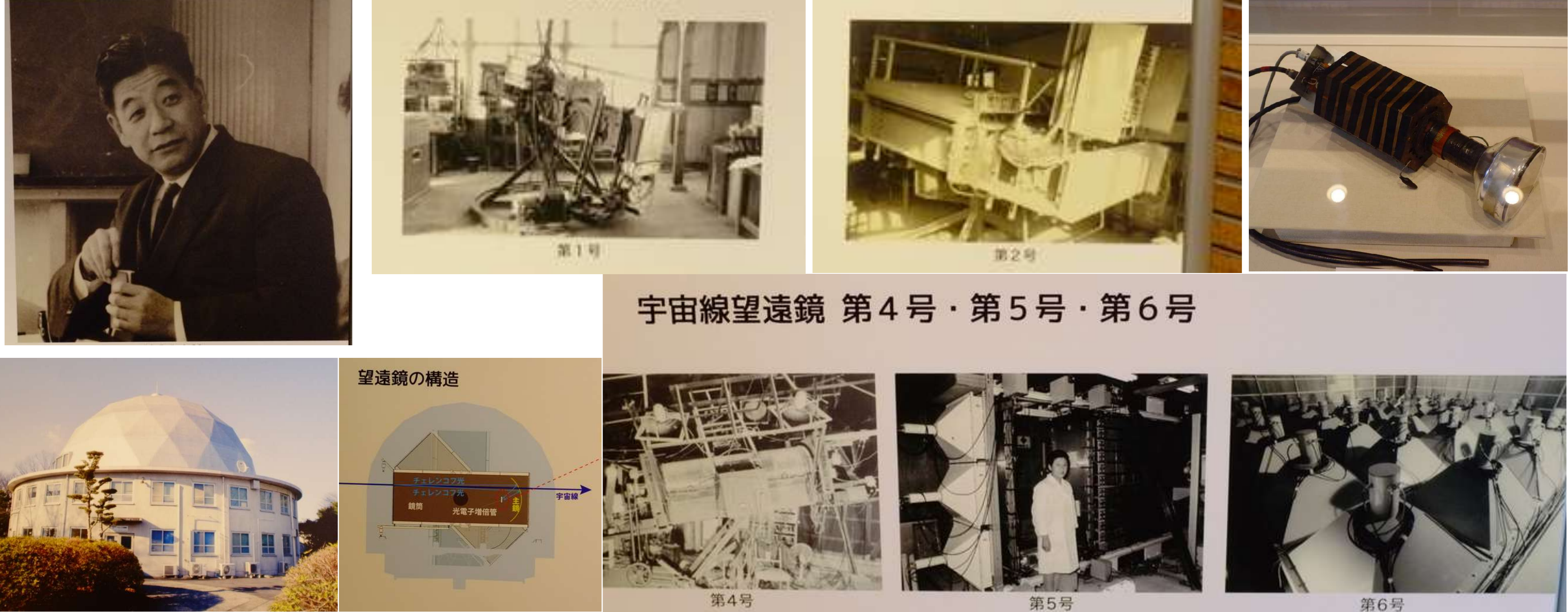}
    \caption{\textbf{Pioneering work performed by Prof.~Yataro Sekido in the development of cosmic ray telescopes.} Motivated by an excess detection in the direction of Orion, Prof.~Yataro Sekido and his colleagues developed a total of six cosmic ray telescopes. Photos taken at Nagoya University Museum.}
    \label{fig:sekido}
\end{figure}

As background to my own research, my interests are observations of ultra-high energy cosmic rays (UHECRs) with energies above 10$^{19}$\,eV (= 10\,EeV) and detector developments. This is because UHECRs are less deflected by Galactic and extragalactic magnetic fields, so their arrival directions are more likely to point back to their sources as shown in Figure~\ref{fig:baton}B\@. I will address the future prospects of ``UHECR astronomy'' and discuss the requirements to clarify the nature and sources of UHECRs.

%\section{Executive overview}
%One of the most intriguing topics is a charged particle astronomy as shown in Figure~\ref{fig:baton}B. 
%Ultra-high energy cosmic rays (UHECRs) are energetic particles with energies above 10$^{19}$\,eV (= 10\,exa-electronvolts, EeV). 
%UHECRs are less deflected by Galactic and extragalactic magnetic fields,
%so their arrival directions are more likely to point back to their sources.
%Figure~\ref{fig:baton}B shows a comparison of energetic cosmic rays and typical cosmic rays,
%as well as possible source candidates in background, such as Centaurus A, M82, M87, neutron star.
%There are many indications for an extragalactic origin of UHECRs:
%6.9$\sigma$ large scale dipole anisotropy above 8\,EeV~\cite{},
%4.0$\sigma$ excess at Centaurus A above 38\,EeV~\cite{},
%2.8$\sigma$ enhancement above 57\,EeV nearby the starburst galaxy M82~\cite{},
%and a 3.3$\sigma$ increase above 25\,EeV in a direction towards the Perseus-Pisces Supercluster~\cite{}.
%Surprisingly, no excess has been found from the Virgo cluster which is the most promising source candidate for UHECRs, this has been dubbed the ``Virgo scandal''.
%Searching for specific sources is more difficult as a result of heavier composition at the highest energies, uncertainties in the Galactic/extragalactic magnetic fields and source density.

\section{Pioneering work of CRI telescope developments in Nagoya}
It is worth mentioning the pioneering work performed by Prof.~Yataro Sekido in CRI telescopes.
In the 1950s, he identified an enhancement of cosmic rays in the direction of Orion~\cite{PhysRev.113.1108}.
Motivated by the result, he and his colleagues constructed a total of six cosmic-ray telescopes as shown in Figure~\ref{fig:sekido}.
There was a special exhibition at Nagoya University Museum to recognize their efforts named ``The voice from the universe -- cosmic ray telescopes of Nagoya University --''\footnote{\url{https://www.num.nagoya-u.ac.jp/english/exhibitions/spot/20230726.html}}.

\section{CRI observatories across the world}
\begin{figure}[b]
    \centering
    \includegraphics[width=1.0\textwidth]{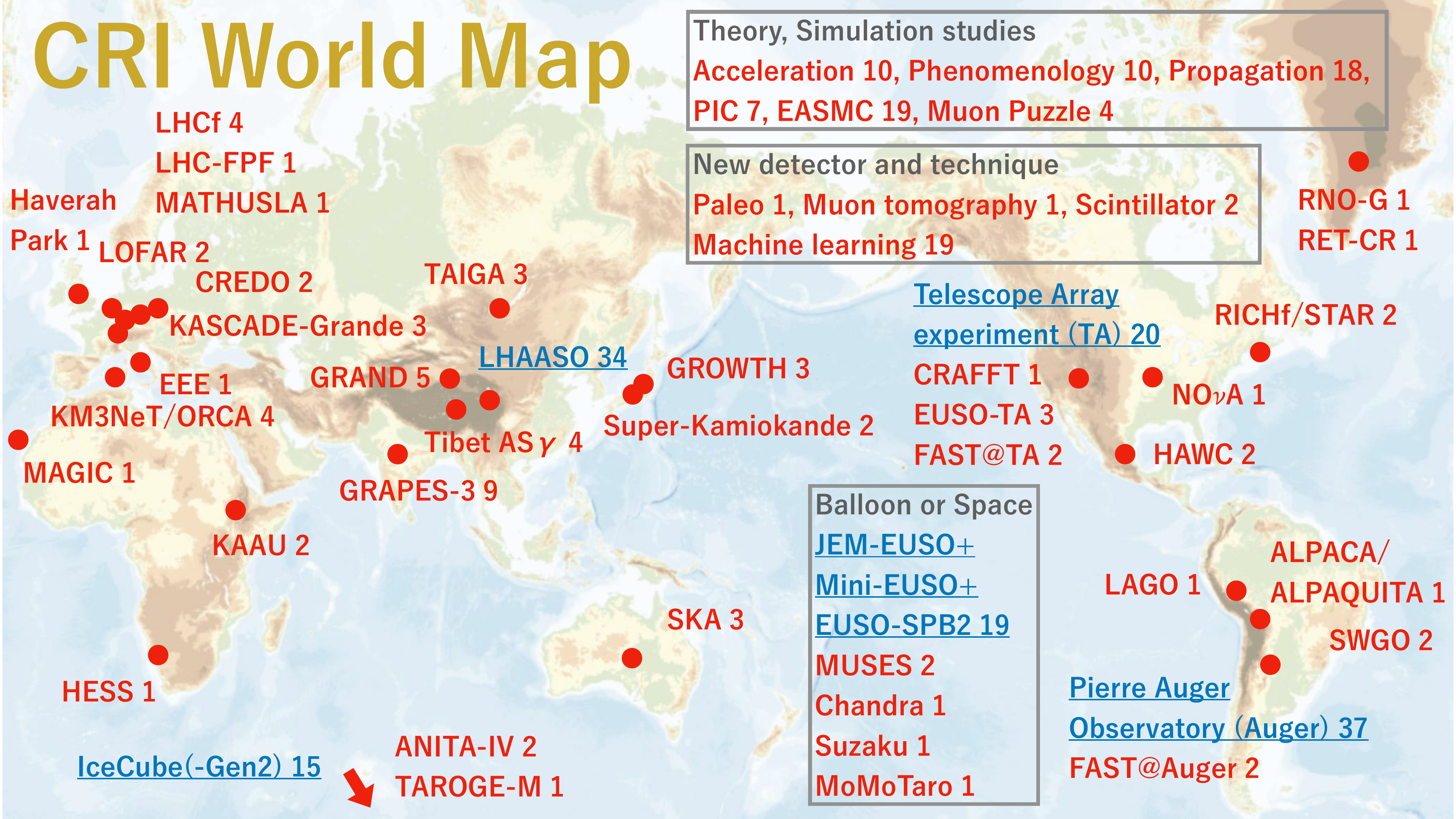}
    \caption{\textbf{CRI World Map.} The filled circles show the locations of CRI experiments around the world.
      The balloon and space experiments are listed in the bottom box. Contributions which were not directly related to any of the listed experiments, such as those focusing on theory, simulations, new detectors and techniques, are listed in the boxes at the top of the figure.
      The number next to each experiment/topic name indicates the number of contributions from that experiment/topic reported in the CRI sessions, categorized by the rapporteur.}
    \label{fig:cri_map}
\end{figure}
In the last 20 years, scientists have built large cosmic ray observatories all over the world. 
Figure~\ref{fig:cri_map} shows a CRI world map indicating locations of the observatories reported in the CRI sessions and their respective number of contributions.
The top-five experiments with the greatest number of contributions were from the Pierre Auger~\cite{AbdulHalim:20232A}, LHAASO~\cite{Wu:2023R5}, Telescope Array~\cite{Kim:2023rW}, JEM-EUSO~\cite{Parizot:2023c4}, and IceCube~\cite{Neilson:2023VJ} collaborations.
% I would like to highlight the idea of installing cosmic ray detectors in schools across Europe, dubbed the ``Extreme Energy Events'' (EEE) experiment~\cite{Noferini:2023OP}. Thus far, there have been no ``Extreme Energy Events'' observed in 4-years of operation.

The Pierre Auger Observatory (Auger) is the world's largest cosmic ray observatory with an effective area of 3000\,km$^{2}$. It is located in Malarg\"{u}e, Argentina and observes the highest energy cosmic rays~\cite{AbdulHalim:20232A}.
The observatory's ongoing upgrade, ``AugerPrime", is in its final commissioning phase and will soon start data-taking with the plastic scintillators and radio detectors which have been installed
on the top of the original water Cherenkov detectors. The upgrade also includes new electronics, higher dynamic range PMTs and underground muon detectors~\cite{AbdulHalim:2023c6,Sato:2023ic}.
The Telescope Array experiment (TA) is the largest cosmic-ray detector in northern hemisphere with an effective area of 700\,km$^{2}$, located in Utah, USA~\cite{Kim:2023rW}\@. It is also currently undergoing an upgrade, called TA$\times4$, to increase the effective area of the array four-fold by installing additional surface detectors~\cite{Kido:2023WK,Fujisue:2023Sx}.
Auger and TA use a hybrid technique to detect extensive air showers, combining a surface detector array (SD) on the ground, overlooked by a fluorescence detector (FD)\@.
The importance of atmospheric monitoring in CRI experiments was summarized in the review talk~\cite{Keilhauer:2023Gx}.

The Joint Exploratory Missions of Extreme Universe Space Observatory (JEM-EUSO) is a mission to observe extensive air showers by placing fluorescence detectors in space~\cite{Parizot:2023c4}.
The Mini-EUSO telescope has been installed onboard the International Space Station, providing measurements of geophysical lightning phenomena~\cite{Marcelli:2023MR}.
The EUSO-SPB2 balloon was launched in May 13, 2023, providing a demonstration of the detector's performance and verification of its design~\cite{Eser:2023Dw,Filippatos:2023P4}.

The Large High Altitude Air Shower Observatory (LHAASO) is an observatory located on Mt.~Haizi in China and detects TeV-PeV gamma rays and charged particles~\cite{Wu:2023R5}.
LHAASO consists of a variety of detectors; a detector array combining scintillation counters and underground muon detectors with 1.3\,km$^2$ coverage (KM2A), a water-Cherenkov detector array with 78,000\,m$^2$ coverage (WCDA), an electron neutron detector array with 1000\,m$^2$ coverage, and a total of 18 wide-field-of-view air Cherenkov telescopes (WFCTA)~\cite{He:2023Hq}\@.
The IceCube observatory is a neutrino observatory with a target volume of 1\,km$^3$ located near the Amundsen-Scott South Pole Station. IceCube possesses a cosmic ray detector which combines an ice-Cherenkov detector array called ``IceTop" and a deep underground muon detector~\cite{Verpoest:2023Fm}. IceTop is now being upgraded to include plastic scintillators which will allow it to be more sensitive to the mass composition of cosmic rays~\cite{Shefali:2023GK}.

I would also like to highlight the idea of installing cosmic ray detectors in schools across Europe, dubbed the ``Extreme Energy Events'' (EEE) experiment~\cite{Noferini:2023OP}. Although thus far there have been no ``Extreme Energy Events'' observed in 4-years of operation, the rapporteur thinks it is a great concept and hopes for successful detection in the near future.

%Thus far, there have been no ``Extreme Energy Events'' observed in 4-years of operation, the rapp

\begin{figure}
    \centering
    \includegraphics[width=1.0\textwidth]{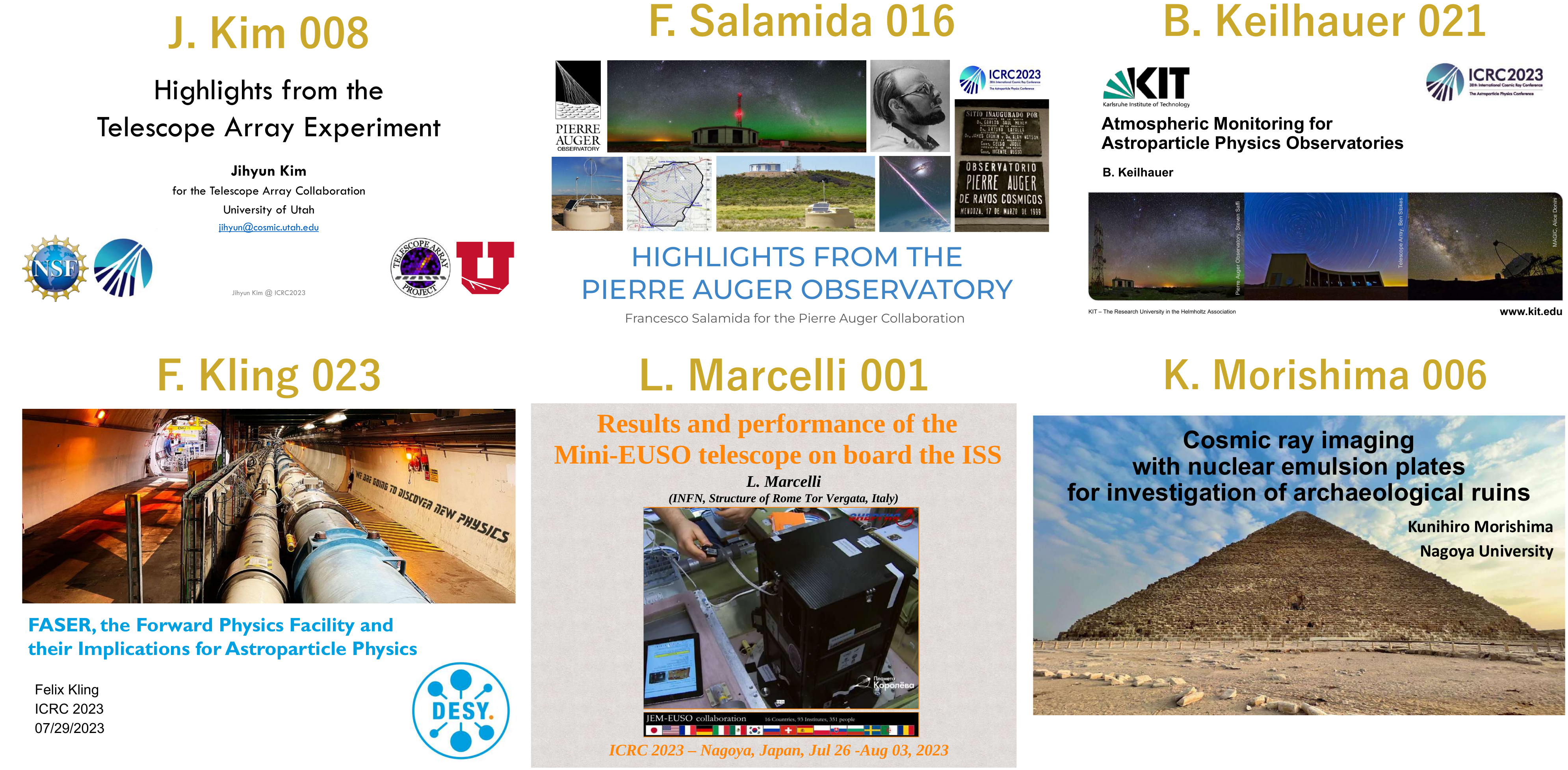}
    \caption{\textbf{Highlight and review talks related to CRI measurements}~\cite{Kim:2023rW,AbdulHalim:20232A,Keilhauer:2023Gx,Kling2023,Marcelli:2023MR,Morishima2023}}
    \label{fig:highlights}
\end{figure}
The highlight and review talks related to CRI measurements are shown in Figure~\ref{fig:highlights}. The review talks from the Forward Physics Facility, which detailed the latest results from hadronic interaction model studies~\cite{Kling2023},
and from the recent progress of cosmic ray applications to non-destructively investigate archaeological ruins (such as Egyptian pyramids)~\cite{Morishima2023} are also shown.

\begin{figure}
    \centering
    \subfigure[\textbf{C}alibration related contributions]{\includegraphics[width=0.99\textwidth]{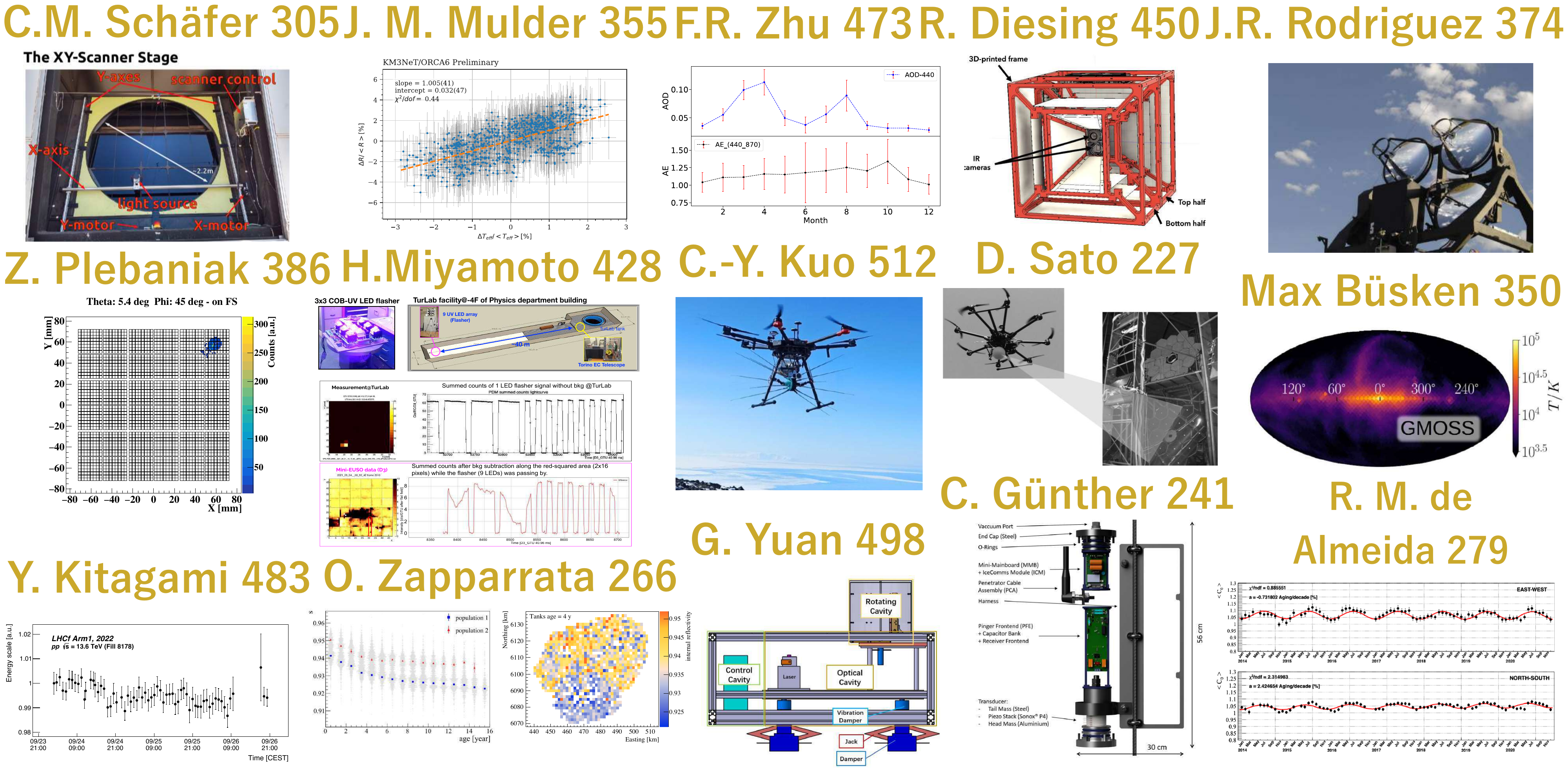}}
    \subfigure[Machine learning related contributions]{\includegraphics[width=0.99\textwidth]{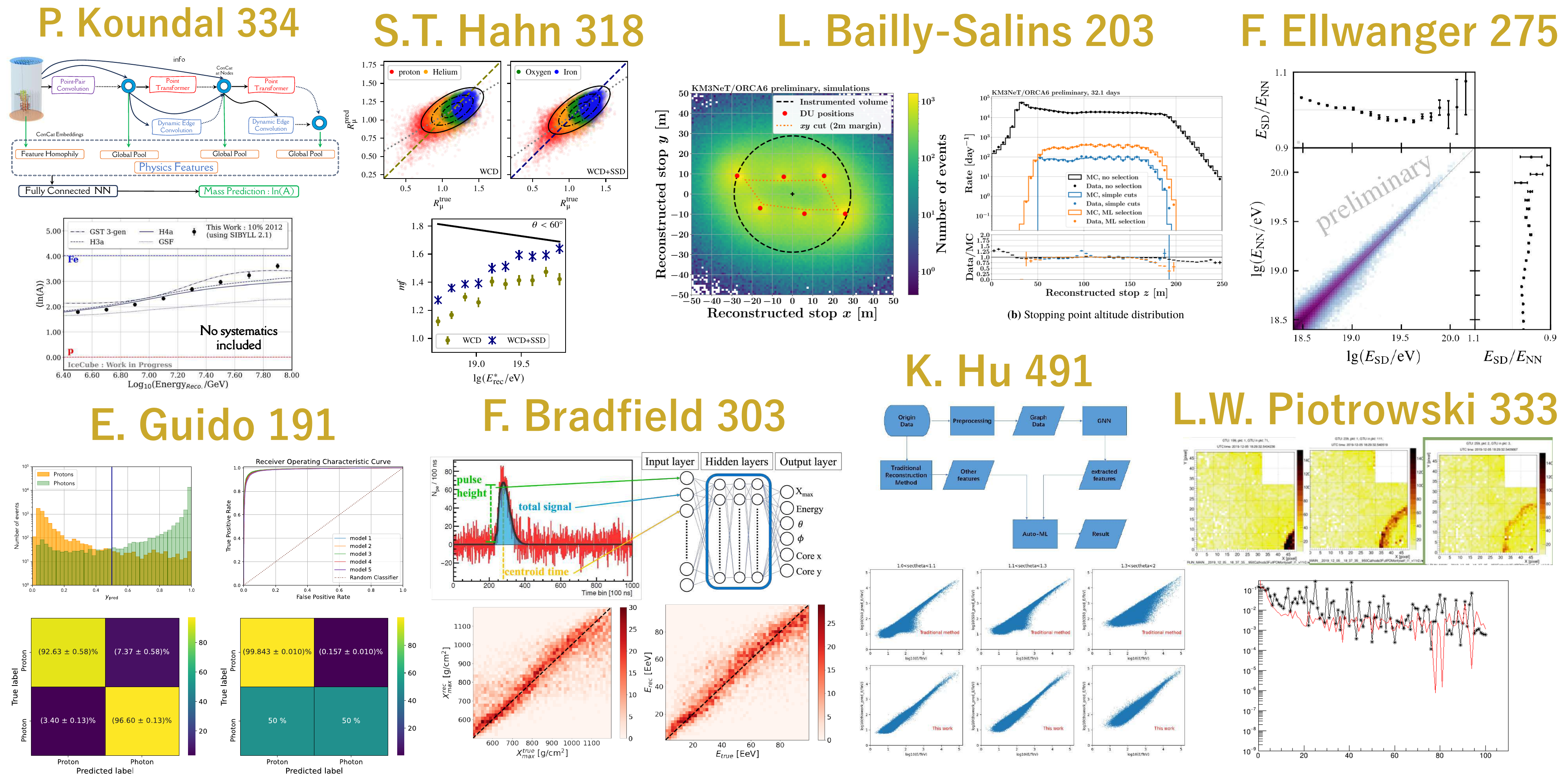}}
    \caption{\textbf{Rapporteur's selection of contributions related to (A) \textbf{C}alibration and to (B) machine learning.} Due to the page limitation, the author and proceedings IDs are indicated as reference.}
    \label{fig:calib_ml}
\end{figure}
\section{Detector calibrations and machine learning techniques}
Focusing on the \textbf{C}alibration of my ``Three Cs'', there were 33 proceedings written detailing calibration methods, instruments and long-term performances. These studies are essential for obtaining accurate final results.
As a new topic, there were 19 contributions regarding machine learning. 
These studies were primarily for the purpose of \textbf{C}ross checking current results and/or for the improvement of current analyses.
They collectively show that utilizing machine learning techniques can provide us new insights into our data. A subset of these contributions, selected by the rapporteur, is highlighted in Figure~\ref{fig:calib_ml}.
Due to the page limitation, not all contributions can be shown. The full list of contributions can be found at \url{https://pos.sissa.it/444/}.

\section{Energy spectrum \textit{-- How frequently do cosmic rays arrive at Earth?}}

\subsection{Electron energy spectrum above TeV}
As an intersection between direct and indirect measurements, the cosmic-ray electron spectrum was measured by imaging atmospheric Cherenkov telescopes, such as MAGIC~\cite{Chai:20239v} and H.E.S.S.~\cite{deNaurois:2023Hx}\@.
The observed spectra indicate a broken power-law structure at 1\,TeV, indicating a softer spectrum above this energy.
Upper limits reported from LHAASO are closer to the extrapolated spectrum of H.E.S.S. at higher energies~\cite{Xiong:2023Mz}.
A discrepancy in the spectrum index above 1\,TeV between MAGIC and H.E.S.S. may be disentangled by future measurements from LHAASO. The rapporteur encourages the organization of an indirect electron working group consisting of the MAGIC, H.E.S.S. and LHAASO \textbf{C}ollaborations.

\subsection{Cosmic ray spectrum around the PeV (``knee'') region}
All particle spectra were reported from HAWC~\cite{Morales-Soto:2023rX,ArteagaVelazquez:20238c}, TAIGA-HISCORE~\cite{Vaidyanathan:2023wx}, GRAPES-3~\cite{Mohanty:2023fZ,Varsi:2023Sc} and Tibet AS$\gamma$~\cite{Takita:20233f,Katayose:2023R1} collaborations.
The individual spectra of proton (= hydrogen), helium and heavier nuclei (atomic number $Z > 3$) were measured by the HAWC experiment~\cite{Morales-Soto:2023rX}, indicating a break feature around 100\,TeV. The maximum energies of each species are proportional to the atomic number $Z$
% corresponding to a charge of primary species
~\cite{ArteagaVelazquez:20238c}.
The spectrum observed by the GRAPES-3 experiment shows a hardening around 100\,TeV~\cite{Varsi:2023Sc}. 
TAIGA-HiSCORE report the spectrum break at 3\,PeV~\cite{Vaidyanathan:2023wx}.
The Tibet AS$\gamma$ experiment estimates a proton-like event abundance in their data, based on simulations using post-LHC interaction models~\cite{Katayose:2023R1}.

The energy spectra in the PeV range were also measured by the LHAASO-KM2A~\cite{Zhang:2023fo,Tian:2023cD}, IceCube~\cite{Rawlins:2023pw}, KASCADE-Grande~\cite{Kang:20231c} and TALE~\cite{AbuZayyad:2023gk} experiments.
LHAASO-KM2A reported the knee feature with high statistics with a break around 3\,PeV~\cite{Zhang:2023fo}\@.
Additionally, they constrained the flux of iron nuclei around PeV energies using large zenith angle showers~\cite{Tian:2023cD}.
IceCube reported the spectrum between PeV and EeV energies. In calculating the energy spectrum, a calibration of the energy scale based on a modulation caused by snow accumulation was essential~\cite{Rawlins:2023pw}.
KASCADE-Grande reported a two component spectrum divided into light (hydrogen + helium + CNO) and heavy (silicon + iron) components~\cite{Kang:20231c}.
TALE measured a spectrum between PeV to EeV with Cherenkov-dominated showers above 2\,PeV~\cite{AbuZayyad:2023gk}, hybrid measurements above 30\,PeV~\cite{Oshima:2023qJ} and by only SD measurements above 100\,PeV~\cite{Komae:2023R9}.

\begin{figure}
    \centering
    \includegraphics[width=1.0\textwidth]{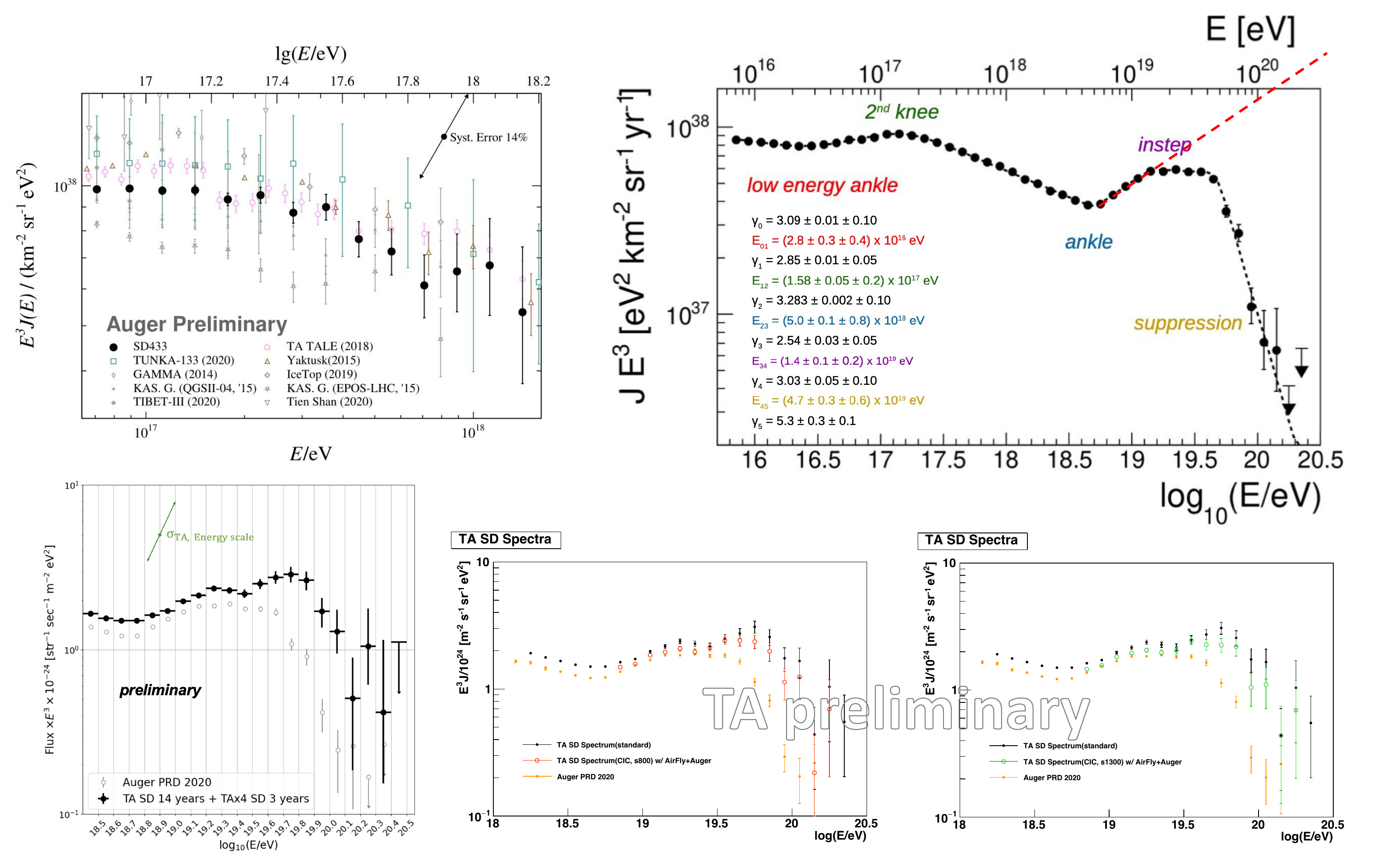}
    \caption{\textbf{Energy spectrum of cosmic rays at the highest energies.} The top figures show the energy spectrum around 100\,PeV measured with the Auger 433\,m array (top-left)~\cite{BrichettoOrquera:202340} and the Auger combined spectrum (top-right)~\cite{AbdulHalim:20232A}. The bottom figures show the TA and TAx4 combined spectrum~\cite{Fujisue:2023Sx} and the TA spectra using different reconstruction methods and physics models for studies of systematic uncertainties~\cite{Ogio:2023S0}.}
    \label{fig:spectrum}
\end{figure}

\subsection{Cosmic ray spectrum above EeV (``ankle'' to ``cutoff'') region}
Together with a precise measurement of the energy spectrum around 100\,PeV using the Auger 433\,m array~\cite{BrichettoOrquera:202340}, 
Auger reported the energy spectrum over a broad energy range from 5\,PeV to beyond 100\,EeV 
%with the largest exposure~\cite{AbdulHalim:20232A} 
as shown in Figure~\ref{fig:spectrum}.
The observed spectrum has a clear softening at 14 EeV before the cutoff, a feature now being referred to as the ``instep''.
TA reported an energy spectrum based on measurements from both TA and TAx4~\cite{Fujisue:2023Sx}.
TA also investigated the change in their energy spectrum results when using the same fluorescence yield model and invisible energy evaluation method used in Auger. Additionally, they investigated changes to the energy spectrum when the estimated signal at a different distance from shower axis was used as the energy estimator~\cite{Ogio:2023S0}.
The hardening of the TA spectrum above 30\,EeV remains even if the same models are assumed as shown in Figure~\ref{fig:spectrum}.
Detailed studies on systematic uncertainties and the energy spectrum in the common declination band are discussed in the report from the joint working group of the Auger and TA \textbf{C}ollaborations~\cite{Tsunesada:2023c0}.

Concerning systematic uncertainties, effects of saturated SDs, optimized distance and lateral density parameterization were investigated to understand possible reasons for differing spectra at the highest energies~\cite{Deligny:2023RB}.
Furthermore, thanks to the data provided by KASCADE and IceTop, the data-driven invisible energy estimation was extended to energies below 100\,PeV~\cite{Vicha:2023px}.
The installation of an Auger water-Cherenkov detector at TA (Auger@TA) has been completed. The detector is currently being prepared for data-taking~\cite{Mayotte:2023WC}.
Continued \textbf{C}ollaboration between Auger and TA is essential to clarify whether differences observed in the northern and southern hemispheres are astrophysical in nature or a result of detector systematics/differences in analysis methods.

\section{Mass composition \textit{-- What kind of particles are cosmic rays?}}
The mass composition of cosmic rays can be estimated from the atmospheric slant depth where an extensive air shower deposits most of its energy, $X_{\max}$. $X_{\max}$ is typically measured with fluorescence detectors. The average value of the $X_{\max}$ distribution at different energies are compared to expectations from Monte Carlo simulations to determine mass fractions.
For surface detector arrays, measurements of the muon component of extensive air showers are needed to estimate the mass composition. This can be achieved through analysis techniques or the installation of underground muon detectors.
\begin{figure}
    \centering
    \includegraphics[width=1.0\textwidth]{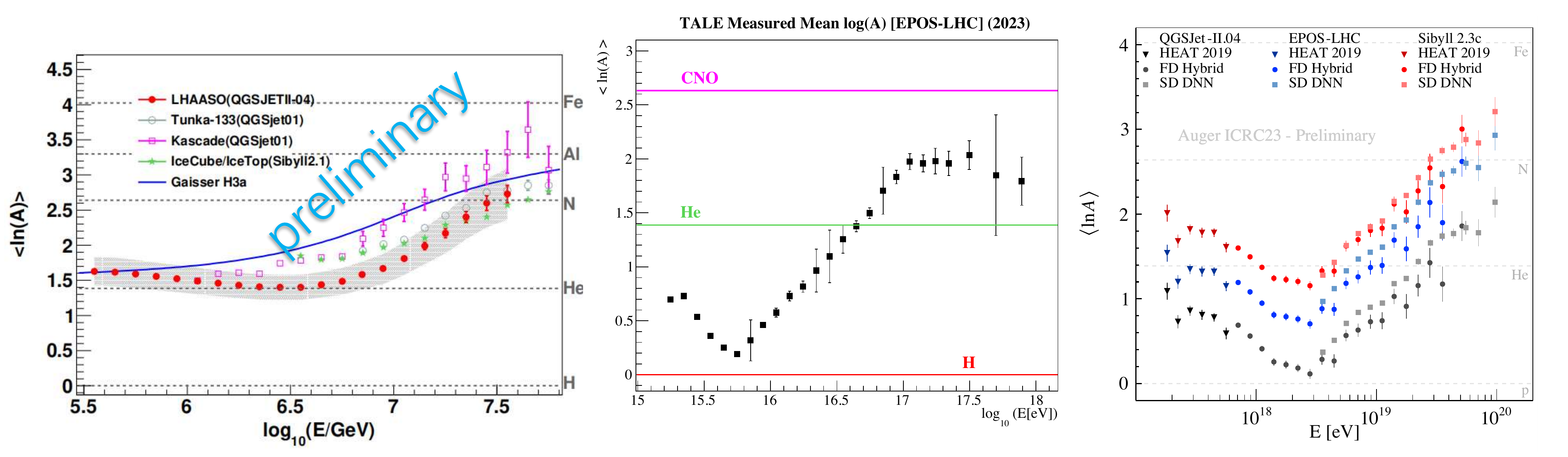}
    \caption{\textbf{The average logarithmic mass numbers.} The figures indicate the mass composition at energies above 3\,PeV measured by LHAASO (left)~\cite{Zhang:2023fo} and TALE (middle)~\cite{AbuZayyad:2023gk}, and above 300\,PeV measured by Auger (right)~\cite{Mayotte:2023Nc}.}
    \label{fig:composition}
\end{figure}

The average logarithmic mass numbers above PeV energies were reported from LHAASO-KM2A~\cite{Zhang:2023fo} and TAIGA-HISCORE~\cite{Vaidyanathan:2023wx}, indicating a dominant helium composition around 3\,PeV, while TALE results indicated a proton composition at 5\,PeV~\cite{AbuZayyad:2023gk} as shown in Figure~\ref{fig:composition}. 
The mass composition measurements using muon components from 10\,PeV to beyond 100\,PeV were reported by IceCube~\cite{Verpoest:2023Fm} and KASCADE-Grande~\cite{ArteagaVelazquez:20231i}, indicating intermediate composition between proton and iron primaries.
Beyond 100\,PeV, TALE hybrid analysis is capable of measuring three mass groups; namely proton, nitrogen and iron. The results indicate a charge-proportional maximum energy for cosmic rays at these energies~\cite{Fujita:2023r8}.
The LOFAR radio detector reported intermediate composition around 300\,PeV based on measurements of $X_{\max}$~\cite{Buitink:2023qO}.

At the highest energies, the latest $X_{\max}$ measurements reported by Auger, for both the FD and SD, indicate a light composition at 3\,EeV followed by a gradual increase in mass number as a function of energy~\cite{AbdulHalim:20232A}. 
Using machine learning techniques, Auger has precisely estimated $X_{\max}$ using only SD measurements. This revealed additional breaks above 3\,EeV , which are in coincidence with the changes in the spectral index in the energy spectrum~\cite{Glombitza:2023}.
The conventional method of $X_{\max}$ determination and the new machine learning technique were compared as a \textbf{C}ross check to validate the machine learning model's performance~\cite{Mayotte:2023Nc} as shown in Figure~\ref{fig:composition}.
Auger reported a ``tension'' between the latest model of QGSJet-II-04 and their observed $X_{\max}$ distributions, and studied how changing the proton-proton cross section and attenuation length affected their reconstructed $X_{\max}$ results~\cite{Tkachenko:2023}.
Using the latest Auger data-set, a mass composition anisotropy at the Galactic plane was reported with a significance of 2.5$\sigma$~\cite{Mayotte:2023Nc}.

\subsection{Studies for systematic uncertainties of mass composition measurements}
A function to describe the profile of an air shower called the ``Greisen function'' was revisited and compared to the ``Gaisser-Hillas'' function which is conventionally used~\cite{Stadelmaier:2023wr}.
Auger modified the form of the Gaisser-Hillas function used in their reconstruction to remove the correlation between parameters of the shower profile~\cite{Bellido:2023Na}.
Atmospheric transparency is one of the most important calibration measurements for fluorescence detectors. Detailed and precise measurements of the daily modulations in atmospheric transparency were studied by Auger, finding systematic uncertainties of $<$4\% in energy and $<$4\,g/cm$^2$ in $X_{\max}$~\cite{Harvey:2023w3}.
The mass composition reported by the Auger and TA working group focused on $X_{\max}$ distributions above 3\,EeV\@. At the current level of statistics and understanding of systematic uncertainties the distributions appear compatible~\cite{Yushkov:2023}.

\subsection{Neutral particle search}
Neutral particles (photons/neutrons) have the advantage of avoiding deflections by the Galactic and extragalactic magnetic fields, and may prove to be the ``smoking gun'' for cosmic ray sources.
A pioneering result was reported from the Auger collaboration using their 433\,m array to constrain the photon flux above 50\,PeV, which in turn gave a constraint on the expected flux of proton-proton interactions in the Galactic halo~\cite{Gonzalez:2023}.
Machine learning techniques for photon searches were adopted by TA, resulting in a constraint on the photon flux above 10\,EeV~\cite{Kharuk:2023T1}.
TA reconstruction method for inclined air showers was studied to increase sensitivities for neutral particles~\cite{Takahashi:2023Hg}.

Although the lifetime of a neutron is only $\sim$900\,seconds, ultra-high energy neutrons can travel a distance of $10 \times (E/(\textrm{EeV}))$\,kpc, where $E$ is the energy of neutrons. Auger reported no observation of excess flux towards the directions of reported Galactic gamma-ray sources, thus providing a constraint on the neutron flux above EeV energies~\cite{Franco:2023}.

\section{Anisotropy \textit{-- Where do cosmic rays come from?}}
\begin{figure}
    \centering
    \includegraphics[width=1.0\textwidth]{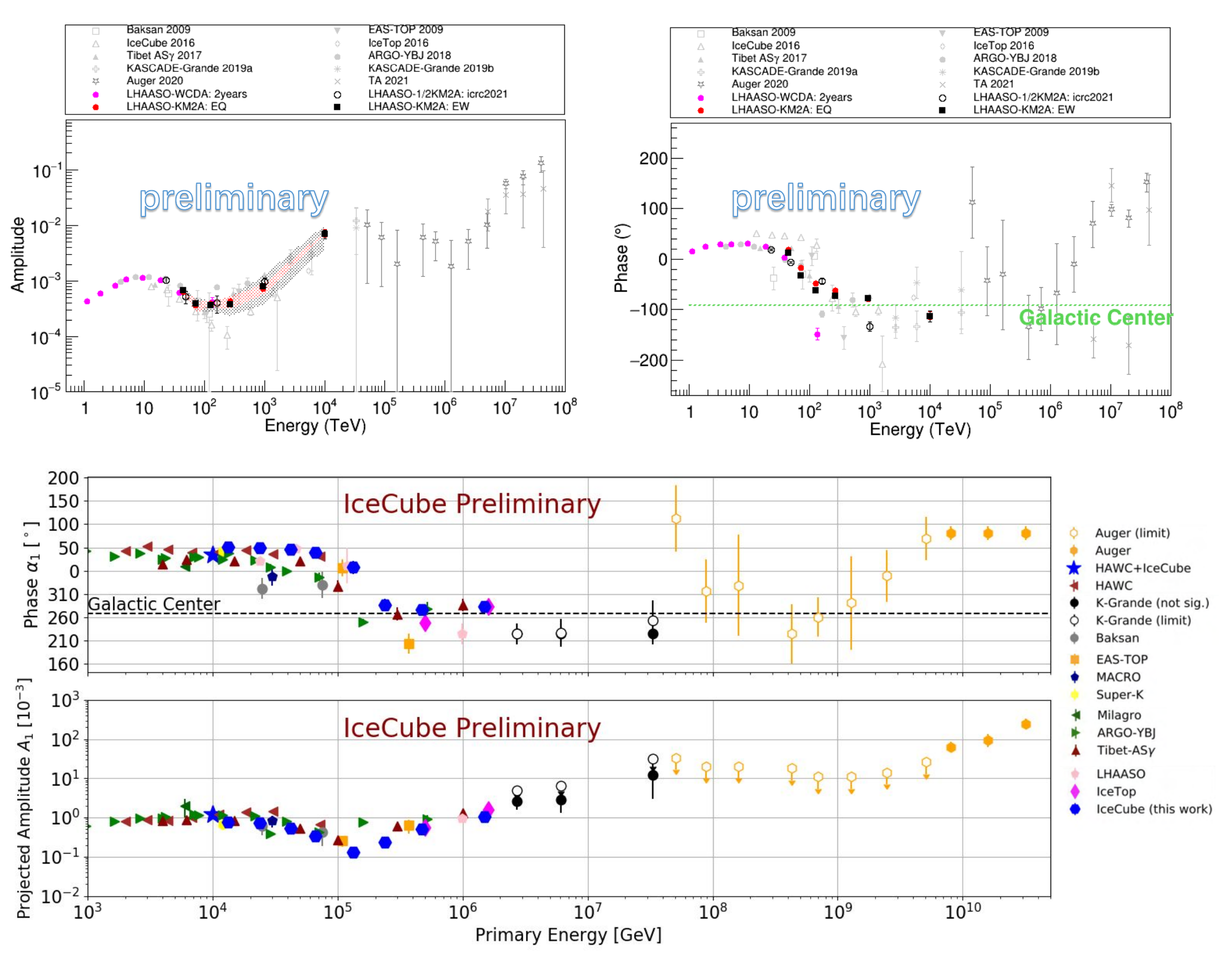}
    \caption{\textbf{Amplitude and phase in large-scale dipole anisotropies of cosmic rays.} The figures show the dipole result reported by LHAASO (top figures)~\cite{Liu:2023/H,Gao:2023E3} and by IceCube (bottom figures)~\cite{McNally:2023Ma}.}
    \label{fig:dipole}
\end{figure}
Anisotropy of cosmic-ray arrival directions is a long-standing and intriguing mystery for cosmic ray researchers.
Since charged particles are deflected by the Galactic and extragalactic magnetic fields, anisotropy searches are sensitive to the strength and structure of these magnetic fields. If the origins of UHECRs are identified, it would be an important breakthrough in astrophysics and astronomy. 
Anisotropy searches are conventionally categorized by small, intermediate and large scales, corresponding to $< 10$\,degrees, $10 - 35$\,degrees and $> 45$\,degrees respectively.

\subsection{Large-scale dipole anisotropy}
The GRAPES-3 experiment studied the small scale anisotropy around 16\,TeV using an angular scale of 10\,degrees. They reported two significant hotspots, region A and B, with significances of 6.8$\sigma$ and 4.7$\sigma$ respectively~\cite{Mohanty:2023fZ,Chakraborty:2023+X}.
The large-scale anisotropies around PeV energies indicated a transition of phase toward 100\,TeV with an increase in the amplitude at energies above 100\,TeV\@. This feature was measured by LHAASO-WCDA and KM2A from 1\,TeV to 10\,PeV~\cite{Liu:2023/H,Gao:2023E3} and measured by IceCube from 10\,TeV to 1\,PeV~\cite{McNally:2023Ma} as shown in Figure~\ref{fig:dipole}.
The feature of a phase transition and amplitude enhancement could be explained by a nearby source model~\cite{Yuan:2023}.
The rapporteur encourages the formation of a working group between the LHAASO and IceCube \textbf{C}ollaborations to disentangle the mystery of the largest cosmic ray accelerators in our galaxy.
The large-scale dipole anisotropy above 8\,EeV was measured by Auger with a significance of 6.9$\sigma$~\cite{AbdulHalim:20232A,Golup:2023}.
The evolution of the dipole amplitude and its direction as a function of energy are consistent with the expectation of a transition from Galactic to extragalactic origins.

\subsection{UHECR ``astronomy''}
\begin{figure}
  \centering
   \subfigure{\includegraphics[width=1.0\textwidth]{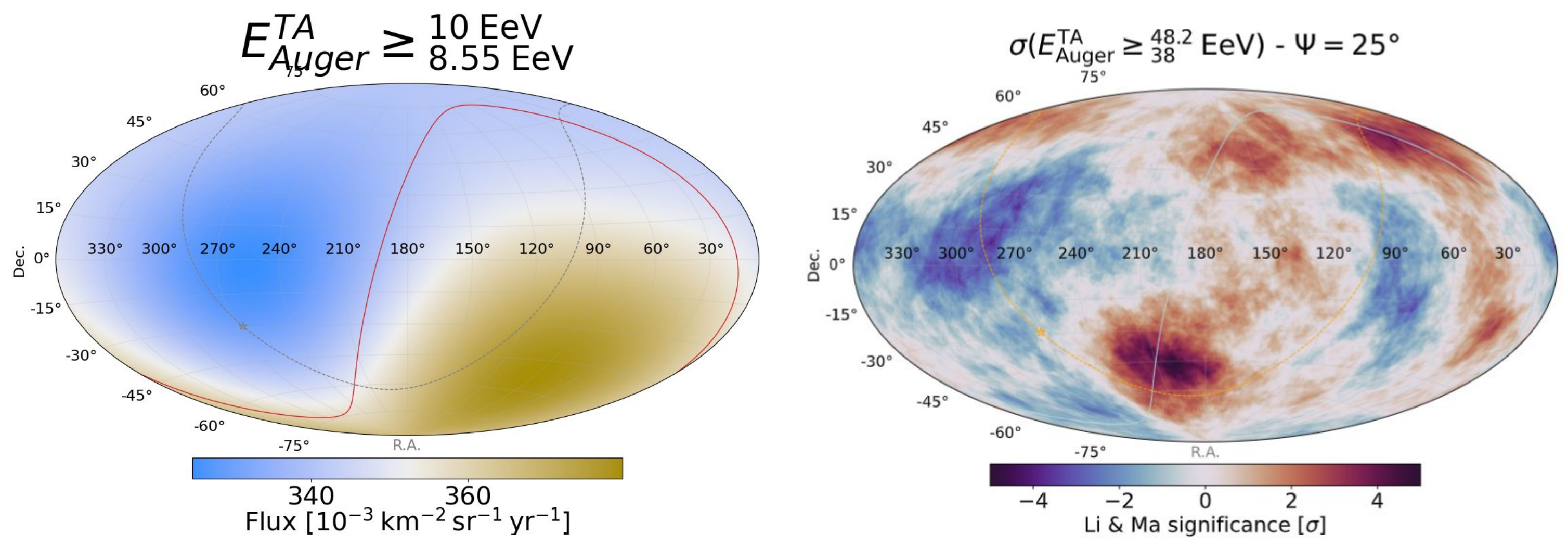}}
   \subfigure{\includegraphics[width=0.75\textwidth]{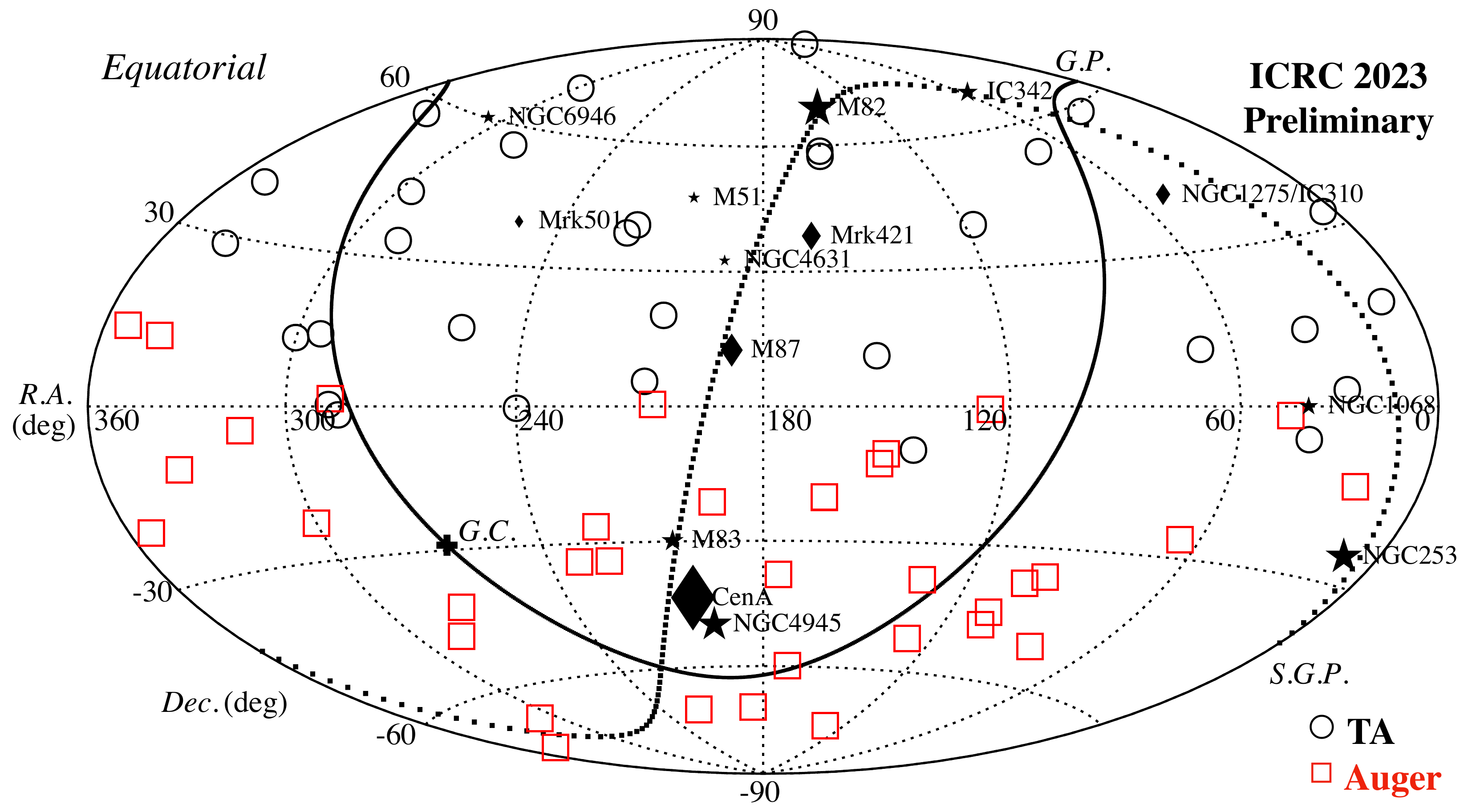}}
    \caption{\textbf{UHECR sky-maps around the ``ankle'', ``cutoff'', and above 100\,EeV.} The flux sky-map of 45$^{\circ}$ oversamplings with energies around the ``ankle'' region (top-left) and significance sky-map of 25$^{\circ}$ oversamplings around the ``cutoff'' region (top-right) in equatorial coordinates reported by the Auger-TA anisotropy working group~\cite{Caccianiga:2023}. The bottom figure indicates the arrival directions of UHECRs above 100\,EeV measured by Auger and TA, together with nearby astronomical source candidates.}
    \label{fig:uhecr_skymap}
\end{figure}

The most significant anisotropy at the highest energies was reported by Auger in the direction of Centaurus A with a significance of 4.0$\sigma$ above 38\,EeV using 27\,degrees oversamplings~\cite{Golup:2023}.
A flux pattern analysis of the southern sky using a catalog of nearby starburst galaxies resulted in a significance of 3.8$\sigma$ under a 9\% anisotropic fraction and 25\,degree angular-scale~\cite{Golup:2023}.
TA shows two hotspots, one of 2.8$\sigma$ above 57\,EeV in the direction of Ursa Major, and of 3.3$\sigma$ above 25\,EeV in the direction of the Perseus-Pisces Supercluster~\cite{Kim:2023rW}.
The TA hotspots were tested by Auger using a compatible exposure. No excesses were found in these directions~\cite{Golup:2023}. 
Further \textbf{C}ross checks and independent measurements are crucial to increase the currently limited statistics and hence reliability of these results.
The \textbf{C}ollaboration between Auger and TA for anisotropy studies was tasked with measuring the all sky-map at the highest energies~\cite{Caccianiga:2023} and making possible interpretations~\cite{Kuznetsov:2023} as shown in Figure~\ref{fig:uhecr_skymap}.
Surprisingly, no excess has been found from the Virgo cluster which is the most promising source candidate for UHECRs. This has been dubbed the ``Virgo scandal''.
%The further data-takings of Auger and TA in both hemispheres and their upgrades are essential to clarify the origins of UHECRs.

Figure~\ref{fig:uhecr_skymap} shows an equatorial sky-map of arrival directions of UHECRs with energies above 100\,EeV observed by Auger in 17-years data set operation~\cite{PierreAuger:2022axr} and TA in 15-years data set.
Although there are intriguing hot/warm spots correlated with nearby possible source candidates around ``cutoff'' energies, there is no apparent correlations/clustering with nearby source candidates above 100\,EeV.
This isotropic distribution was not foreseen 20 years ago and is likely due to a heavier composition at the highest energies and uncertainties in the Galactic/extragalactic magnetic fields and source density.
Further data-taking by Auger and TA in both hemispheres and their upgrades are essential to clarify the origins of UHECRs and to establish ``UHECR astronomy''.

\subsection{Source constraints using spectrum, composition, and anisotropy}
The spectrum, composition, and anisotropy of cosmic rays should be linked to their nature and origin.
LHAASO reported these three observables around the knee region (3\,PeV) showing a spectral break,
transition to a heavier composition and increase in amplitude of dipole anisotropies,
indicating a maximum energy to which cosmic rays are accelerated by Galactic sources~\cite{Zhang:2023fo,Gao:2023E3}.
Auger reported these three observable at the highest energies~\cite{AbdulHalim:20232A}. The spectral features of the ankle, instep and suppression are in coincidence with the ``breaks'' of the elongation rate of $X_{\max}$~\cite{Glombitza:2023}.
The dipole amplitude was also observed to increase above 3\,EeV, with a shift in phase towards a direction away from the Galactic center, supporting an extragalactic origin~\cite{Golup:2023}.
Combining results of the spectrum, composition and anisotropy by Auger, a source model of the gamma-ray emitted active galactic nuclei was disfavored assuming the cosmic ray flux is proportional to the gamma-ray flux of sources~\cite{Bister:2023,PierreAuger:2022atd}.
The rapporteur expects future analyses combining all three observables to shed light on the origin and nature of UHECRs.

\section{Hadronic interaction models \textit{-- How do high-energy particles interact?}}
\subsection{``Muon puzzle'': Discrepancy in muon number between data and simulations}
%FRASER ONLY
The muon number is a key piece of information in 
% understanding hadronic interaction models.
determining the accuracy of hadronic interaction models.
IceCube and IceTop reported a discrepancy between the muon numbers estimated from the number of high energy muons detected above 500\,GeV and the muon densities measured at 600\,m and 800\,m in the latest interaction models~\cite{Verpoest:2023Fm}.
The Tibet AS$\gamma$ experiment studied the muon numbers with a tension of Sibyll models for large shower size~\cite{Haung:2023}.
% ?????? experiment found a discrepancy between the muon number for large showers as estimated from measurements and with SIBYLL models (?)
Surprisingly there was a re-analysis of the Haverah Park experiment's data. The analysis found no significant discrepancy in muon numbers estimated from data and simulations~\cite{Cazon:2023ik}.
The neutrino experiment KM3Net, located in the 
% deepest seas of the 
Mediterranean sea, also showed a muon number of 1.4 -- 1.8 times larger than MC expectations~\cite{Romanov:2023Z9}.
KASCADE-Grande re-analyzed their data using the latest models and found an intermediate composition between proton and iron primaries, intriguingly showing ``no muon puzzle''~\cite{ArteagaVelazquez:20231i}.
Overall, these results indicate a softer muon spectrum than models in simulations; fewer high energy muons and more low energy muons.

\begin{figure}
    \centering
    \includegraphics[width=1.0\textwidth]{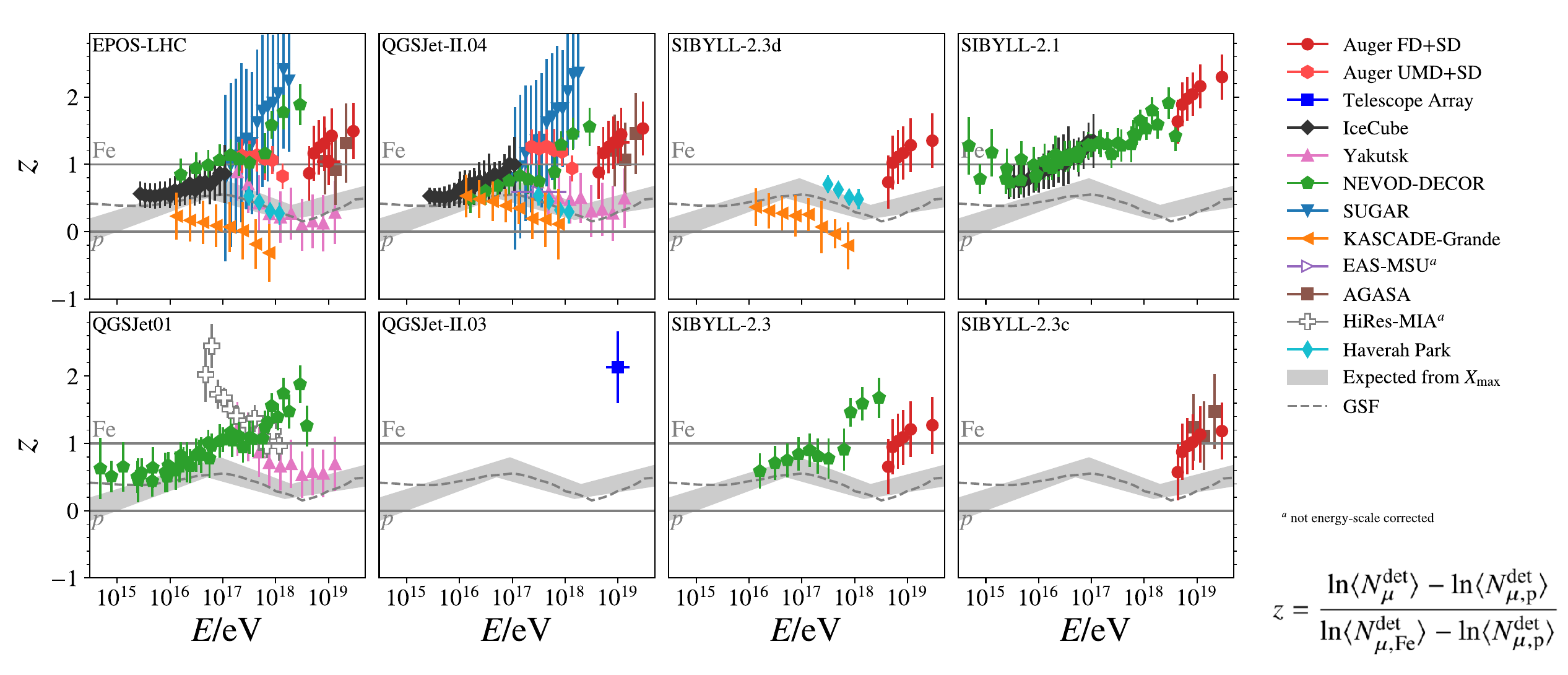}
    \caption{\textbf{Muon densities of extensive air showers measured by a variety of experiments above 1\,PeV} The $Z$ parameter (y-axis) indicates the relative difference in muon number between data and a pure proton composition compared to that between pure iron and pure proton (as estimated by simulations). 
    % The measured results are compared to predictions of different hadronic models
    Estimates of $Z$ using a variety of hadronic interaction models are shown~\cite{ArteagaVelazquez:20236d}.}
    \label{fig:muon}
\end{figure}

% Auger reports a muon puzzle from the main array and also underground muon detectors~\cite{AbdulHalim:20232A},
% and also the model independent measurement of ``composition mixture'' by muon number and $X_{\max}$~\cite{Stadelmaier:2023NZ}.
% The results shows the composition is not pure composition around 10\,EeV and above 20\,EeV,
% consistent with the observed FD $X_{\max}$ values.
% Figure~\ref{fig:muon} shows muon densities above 5\,EeV reported in the working group on hadronic interactions and shower physics (WHISP)~\cite{ArteagaVelazquez:20236d}.
Auger observes a muon deficiency in comparison to simulations from both their main array and underground muon detectors~\cite{AbdulHalim:20232A}. They also provide an independent measurement of ``composition mixture'' using muon number and $X_{\max}$~\cite{Stadelmaier:2023NZ}. The results show the composition is not pure around 10\,EeV and above 20\,EeV, consistent with the observed FD $X_{\max}$ values. Figure~\ref{fig:muon} shows muon densities above 1\,PeV reported by the working group on hadronic interactions and shower physics (WHISP)~\cite{ArteagaVelazquez:20236d}.

% Changed here to "collider experiments"
\subsection{Zero-degree measurements at collider experiments}
Forward neutral particles are measured by LHCf, with the results being jointly analyzed in \textbf{C}ollaboration with ATLAS~\cite{Tiberio:2023of}\@.
The LHCf measures neutrons, the $\eta$ meson production rate and the $\eta/\pi^0$ ratio to tune the interaction models~\cite{Piparo:2023PB}.
The collision of protons and oxygen nuclei is scheduled for 2024~\cite{Tiberio:2023of}.
There have also been successful and promising results from FASER\@. The Forward Physics Facility plans to measure forward going high energy TeV neutrinos to constrain hadronic interaction models~\cite{Soldin:2023hZ}.

There are plans to upgrade the Sibyll and EPOS-LHC interaction models to Sibyll$^{\star}$~\cite{Riehn:2023Wf} and EPOS-LHC-R~\cite{Pierog:2023OB} respectively. The ``MOCHI'' parameterization will allow for the study of effects of ad-hoc modifications to the cross-section, multiplicity, and elasticity parameters~\cite{Ebr:2023dR}.
Currently, no model can reproduce muon observables in all energy ranges, thus \textbf{C}ross checks between theoretical and experimental results as well as further \textbf{C}ollaboration is required for fine-tuning the models.

\subsection{Simulations for extensive air showers}
The extensive air shower simulation software CORSIKA is now being upgraded to CORSIKA 8 and will be written in C++~\cite{Huege:2023,Albrecht:20238R}.
Radio emission has been implemented in CORSIKA 8 and is ready to use~\cite{AlvesJr:2023Hz,Alameddine:2023Sp}.
COSMOS X is an independently developed piece of software which also simulates extensive air showers and will be important for \textbf{C}ross checks~\cite{Sako:2023Ox}. 
A user friendly package named ``Chromo" is being prepared and will require the user to only write a few lines of code to simulate particle interactions.
Cherenkov light emission packages are being developed which utilize GPUs~\cite{Baack:2023} and python (CHASM)~\cite{Buckland:2023YC}\@.
For the purpose of understanding inclined, high energy neutrino induced air showers, simulation studies involving 100 PeV extensive air showers in the upper atmosphere~\cite{Krizmanic:2023O0}, atmospheric skimming showers~\cite{Tueros:2023rx}, and radio emission from inclined showers~\cite{Chiche:2023oN} were reported. 
A study of the neutron component of extensive air showers was also performed~\cite{Schimassek:2023I6}.

\section{Geophysics and interdisciplinary research \textit{-- How useful are cosmic rays for us?}}
Thunderstorms and lightning are trendy topics in the intersection of geophysics and cosmic ray physics.
LHAASO-KM2A measured a correlation between the strength of the electric fields of thunderstorms and cosmic ray shower rates~\cite{Zhou:2023Vq}.
TA recorded a lightning strike using both a high-speed camera and cosmic ray detector~\cite{Abbasi:2023Js}. 
Auger reported sub-millisecond pulses of gamma-rays measured by their surface detectors~\cite{Colalillo:2023}.
Also the ELVES and halo which are atmospheric transient emissions related to lightnings were precisely measured by their fluorescence detector~\cite{Mussa:2023}.

The GROWTH project is a new initiative to deploy portable gamma-ray detectors across Japan's Kanagawa prefecture~\cite{Nakazawa:2023wo}.
GROWTH reported a possible connection between a cosmic ray interaction and a triggering of the lightning flash in thundercloud~\cite{Tsurumi:2023Zd}, and measurements of the gamma-ray glow spectrum~\cite{Wada:2023Y+}. The results were consistent with the expectation from bremsstrahlung emission~\cite{SousaDiniz:2023O3}. This detector can be used for exploring the water resources at the Moon~\cite{Tsuji:2023o}.
In space, Mini-EUSO showed results measuring multiple ring ELVES~\cite{Marcelli:2023MR,Romoli:2023C0}.
A seasonal variation in the number of lightning strikes was reported by GRAPES-3~\cite{Nayak:2023J3}\@.
LOFAR has constructed a map of lightning strikes~\cite{Mulrey:2023v8}.
On a geophysics note, the Hunga Tonga-Hunga Ha'apai volcano eruption was detected by GRAPES-3~\cite{Hariharan:2023Gw} and HAWC~\cite{Sandoval:20231J}.

There was significant progress in investigating large-scale historical objects using the cosmic ray imaging technique of ``muography''.
The void room and north face corridor of the Pyramid of Khufu were revealed by a nuclear emulsion detector~\cite{Morishima2023}.
The north face corridor was confirmed by a photograph using a fiber-scope~\cite{Su:2023IJ}.
The rapporteur believes a portable detector with directional sensitivity to muons is important to achieve precise measurements and progress muography.

\section{Theory and phenomenology \textit{-- How to interpret the experimental results?} }
Coming up with acceleration mechanisms which could accelerate cosmic rays to the highest energies remains challenging and is a topic under debate.
The theory and phenomenology of the sites of acceleration and the effects of propagation on UHECRs are crucial for interpreting experimental results.

Star clusters and shocked stellar winds were proposed as possible cosmic ray sources to explain both the energy spectrum and the transition in mass composition between PeV to EeV energies~\cite{Bhadra:2023Cu}.
Galaxy clusters~\cite{Simeon:20233q}, %%\cite{Boula:2023DY}
ultra-fast outflows~\cite{Peretti:2023aO}, stratified jets of active galactic nuclei~\cite{Wang:2023ab} and jet back-flows~\cite{Araudo:2023ie} were considered to be the highest energy acceleration mechanisms.
Observational constraints on transient scenarios, such as long gamma-ray bursts with low luminosities and tidal disruption events were studied~\cite{Condorelli:2023z/}.
A particle-in-cell (PIC) simulation revealed that heavy ions are accelerated efficiently because of their larger mass-to-charge ratio~\cite{Tomita:2023A0}.
3-dimensional Magneto-Hydro-Dynamic (MHD) and test particle simulations were performed to investigate the particle acceleration and turbulent field amplification in a highly relativistic shock for the first time~\cite{Morikawa:2023te}. 
In the non-relativistic system, the efficient ion acceleration at a perpendicular shock, i.e. the angle between the background magnetic field and the shock normal is 90 degrees, was demonstrated by 3-dimensional hybrid (kinetic ions---fluid electrons) simulations~\cite{Orusa:20231g}.
Developing numerical techniques~\cite{Fulat:2023R+,Jikei:2023} and implementing machine learning techniques~\cite{TorralbaPaz:2023y7} in PIC simulations allowed us to investigate the electron acceleration and the magnetic field amplification in the non-relativistic shocks.
%The PIC simulation of non-relativistic shocks in turbulence~\cite{Fulat:2023R+} and the amplification of magnetic fields by 3D PIC~\cite{Jikei:2023} were calculated to understand the mechanism of shock acceleration.

The difference in energy spectra between Auger and TA at the highest energies was interpreted as a contribution from a local source~\cite{Plotko:20237z}.
Differences in the mass composition between the two experiments could also be attributed to such a source~\cite{Ryu:2023}.
% The mass composition difference could be interpreted as being due to such a source~\cite{Ryu:2023}.
However, it is difficult for these local source models to explain the isotropic distribution reported at the highest energies.
%%However, these local source models have to overcome difficulties to explain the isotropic distribution reported at the highest energies.
%As M82 is not a promising source as the highest energy accelerator, the TA hotspot could be created from the ``echo'' of Centaurus A active past~\cite{Taylor:2023RP}.
As M82's promise as a source for accelerating the highest energy cosmic rays has somewhat diminished, an alternative explanation for the origin of the TA hotspot being an ``echo'' of Centaurus A's active past was proposed~\cite{Taylor:2023RP}.
Possible source models where the origin of the TA hotspot is M82~\cite{Watanabe:2023kX} and/or M83~\cite{Bourriche:2023N/} were also suggested.
% ?????
% 
Taking into account the maximum rigidity diversity of sources, it was found that, universally, there is a maximum energy that can be reached by these accelerators~\cite{Oikonomou:2023Ya}.

An advantage of charged particles is that measuring deflections from sources allows the strength and structure of the Galactic and extragalactic magnetic fields to be inferred.
% The observed large-scale dipole anisotropy above 8\,EeV was attempted to reproduce the dark matter distribution and deflections in the Galactic and extragalactic magnetic fields~\cite{Farrar:2023vB}. 
An attempt to reproduce the observed large-scale dipole anisotropy above 8\,EeV using the observed distribution of dark matter and the Galactic and extragalactic magnetic fields was performed~\cite{Farrar:2023vB}.
A new model of the coherent Galactic magnetic field which includes the final polarized intensity maps from WMAP and Planck was presented~\cite{Unger:2023Rn}.
The expected excess distribution of UHECRs was estimated by considering the propagation of UHECRs in a turbulent intergalactic magnetic field~\cite{Dolgikh:2023tq}. 
Effects of the Galactic magnetic field on energy spectrum and mass composition are investigated~\cite{Higuchi:2023mL}.
Assuming an individual extremely high-energy cosmic ray of a specific primary species, a method to distinguish between steady and transient or highly variable sources, accounting for deflections by the Galactic and extragalactic magnetic fields, was reported as a ``treasure map''~\cite{Globus:2023AK}.

\section{Developments in next-generation CRI observatories}
% It is essential to build next generation observatories for unraveling sources of UHECRs.
To clarify sources of UHECRs, next generation observatories with extremely large exposures are required.
One method of obtaining such large exposures is by using satellites. The satellite experiments POEMMA-Balloon and Radio (PBR)~\cite{Olinto:2023ZQ} and MUSES~\cite{Aloisio:2023K6,Bertaina:2023P6} 
are planning to launch in 2026.
The uniform exposure in the northern and southern hemispheres is important for \textbf{C}ross checks to confirm the reported hints of anisotropies at the highest energies.

Although primarily focused on radio observations of celestial objects, Square Kilometer Array (SKA) will have the ability to measure extensive air showers with high resolution; $<$8~g/cm$^2$ in $X_{\max}$ and 3\% in energy~\cite{Buitink:2023qO}.
Similarly, LOFAR 2.0 with low (30-80 MHz) and high (120-240 MHz) band antennas will also allow for studies of the radio emission from extensive air showers~\cite{Mulrey:2023v8}.
Radio arrays specifically designed to measure extensive air showers from neutrinos and cosmic rays are also being developed. Formulating a robust internal trigger for such arrays is a challenging but essential task.
RNO-G reported successful measurements of cosmic rays using their internal trigger~\cite{Henrichs:2023XE}. 
GRAND has tested and validated their internal trigger in the laboratory and will progress to testing in the field~\cite{LeCoz:2023yz,Mitra:2023}.

The IceCube Surface Array Enhancement (SAE) is an installation of plastic scintillators at IceTop which, when combined with the current ice-Cherenkov detector, will increase sensitivity to the mass composition of primary cosmic rays~\cite{Shefali:2023GK}.
SAE prototypes have been installed at both Auger and TA sites for field measurements and \textbf{C}ross checks.
IceCube-Gen2 is a powerful cosmic-ray detector and covers a broad energy range from 100\,TeV to 10\,EeV.
Combined measurements of $X_{\max}$ and the muon component of air showers will provide a high precision measurement of the mass composition in this energy range~\cite{Coleman:2023mP}.
A prototype of ALPACA, ALPAQUITA, combining plastic scintillators and underground water Cherenkov detectors in Bolivia, has started data-taking to search for Galactic PeVatrons in the southern hemisphere~\cite{Kawata:2023mA}. 
The moon shadow was observed with a significance of 6.9$\sigma$, demonstrating the detector's performance.

To achieve an unprecedented exposure from ground based methods, cost-effective fluorescence detectors are being developed. 
FAST~\cite{Sakurai:20236q} is utilizing a simplified mirror setup/optics, 
whilst CRAFFT~\cite{Tameda:2023vg} is using Fresnel lens optics. 
FAST prototypes have been installed at both Auger and TA for \textbf{C}ross checks on energy and $X_{\max}$ scales~\cite{Sakurai:20236q}.
The concept of a Global Cosmic Ray Observatory (GCOS) poses a promising science case for high energy physics, fundamental physics, particle physics and solar, geo and atmospheric physics~\cite{AlvesBatista:20233O}.
The future objectives of UHECR science, outlined in the Snowmass paper~\cite{Coleman:2022abf}, were reported in the contributions~\cite{Sarazin:2023Cz,Schroeder:2023uc}.
%The future science objectives of UHECRs were reported~\cite{Sarazin:2023Cz,Schroeder:2023uc} and also summarized in the Snowmass paper~\cite{Coleman:2022abf}.
The World-one \textbf{C}ollaboration is absolutely essential for the timely realization of a future observatory.

\section{Summary and future perspectives}
The origin and nature of UHECRs are still inconclusive as of ICRC2023.
Looking back 20 years, scientists and researchers have been successful in constructing giant ground based observatories and pioneering measurements from space, resulting in a significant improvement in UHECR detection.
Unfortunately the origin and nature of UHECRs have proven to be more complicated than our original expectations.
The isotropic distribution of UHECRs implies a heavier composition at the highest energies and uncertainties in the Galactic/extragalactic magnetic fields and source density. 
Interdisciplinary studies such as combining geophysics and comic-ray applications have made remarkable progress.

In ICRC2023 the rapporteur was delighted to meet the enthusiastic next-generation of cosmic ray scientists and discuss thought-provoking ideas and promising future projects.
Hopefully the proceedings of ICRC\textbf{2043}, possibly held in Japan, will be described as follows;
``After decades of attempts to discover the origin of ultra-high energy cosmic rays, we have established a new astronomy with ultra-high energy charged particles, firmly confirming their origin and nature''.

\section*{Acknowledgements}
I would like to thank (in alphabetical order) ALPACA, ANITA, CORSIKA, CRAFFT, CREDO, EEE, FAST, GCOS, GRAND, GRAPES-3, GROWTH, HAWC, H.E.S.S., IceCube, JEM-EUSO, KASCADE-Grande, KM3Net, LAGO, LHAASO, LHCf, LOFAR, MAGIC, NO$\nu$A, NUSES, Pierre Auger, POEMMA, RHICf, RNO-G, SKA, Super-Kamiokande, SWGO, TAIGA, Telescope Array and Tibet AS$\gamma$ collaborations for presenting their latest results in the CRI sessions.
I would like to express special thanks to Foteini Oikonomou and Sara Tomita for valuable discussions regarding theory and phenomenology.
I would like to thank Fraser Bradfield for carefully reading and polishing the proceedings.
Finally I deeply appreciate all of the scientists and students who I met and had discussions with at ICRC2023.
Without your cooperation, this proceedings would not have been possible.
I am looking forward to seeing you again at upcoming ICRCs.

\fontsize{10pt}{9pt}\selectfont
\bibliography{main}

\providecommand{\href}[2]{#2}\begingroup\raggedright\begin{thebibliography}{100}

\bibitem{Olinto:2004hc}
A.~V. Olinto, {\it {Rapporteur talk for ultrahigh energy cosmic rays (HE 1.3,
  1.4, 1.5): Messengers of the extreme universe}},  in {\em {28th International
  Cosmic Ray Conference}}, pp.~299--319, 4, 2004.
\newblock \href{http://arxiv.org/abs/astro-ph/0404114}{{\tt astro-ph/0404114}}.

\bibitem{PhysRev.113.1108}
Y.~Sekido, S.~Yoshida, and Y.~Kamiya, {\it Point source of cosmic rays in
  orion},  {\em Phys. Rev.} {\bf 113} (Feb, 1959) 1108--1114.

\bibitem{AbdulHalim:20232A}
{\bf Pierre Auger} Collaboration, F.~Salamida, {\it {Highlights from the Pierre
  Auger Observatory}},  {\em PoS} {\bf ICRC2023} (2023) 016.

\bibitem{Wu:2023R5}
{\bf LHAASO} Collaboration, S.~Wu and S.~Chen, {\it {Highlight of LHAASO
  science results on PeVatrons}},  {\em PoS} {\bf ICRC2023} (2023) 010.

\bibitem{Kim:2023rW}
{\bf Telescope Array} Collaboration, J.~Kim, {\it {Highlights from the
  Telescope Array Experiment}},  {\em PoS} {\bf ICRC2023} (2023) 008.

\bibitem{Parizot:2023c4}
{\bf JEM-EUSO} Collaboration, E.~Parizot and M.~Casolino, {\it {Overview of the
  JEM-EUSO program for the study of ultra-high-energy cosmic-rays from space}},
   {\em PoS} {\bf ICRC2023} (2023) 208.

\bibitem{Neilson:2023VJ}
{\bf IceCube} Collaboration, N.~K. Neilson, {\it {Highlights from the IceCube
  Neutrino Observatory}},  {\em PoS} {\bf ICRC2023} (2023) 017.

\bibitem{AbdulHalim:2023c6}
{\bf Pierre Auger} Collaboration, F.~Convenga, {\it {The performances of the
  upgraded surface detector stations of AugerPrime}},  {\em PoS} {\bf ICRC2023}
  (2023) 392.

\bibitem{Sato:2023ic}
{\bf Pierre Auger} Collaboration, R.~Sato, {\it {AugerPrime implementation in
  the DAQ systems of the Pierre Auger Observatory}},  {\em PoS} {\bf ICRC2023}
  (2023) 373.

\bibitem{Kido:2023WK}
{\bf Telescope Array} Collaboration, E.~Kido, {\it {Updates of the surface
  detector array of the TAx4 experiment}},  {\em PoS} {\bf ICRC2023} (2023)
  239.

\bibitem{Fujisue:2023Sx}
{\bf Telescope Array} Collaboration, K.~Fujisue, {\it {Measurement of the
  cosmic ray energy spectrum with the TAx4 SD array}},  {\em PoS} {\bf
  ICRC2023} (2023) 308.

\bibitem{Keilhauer:2023Gx}
B.~Keilhauer, {\it {Atmospheric Monitoring for Astroparticle Physics
  Observatories}},  {\em PoS} {\bf ICRC2023} (2023) 021.

\bibitem{Marcelli:2023MR}
{\bf JEM-EUSO} Collaboration, L.~Marcelli, {\it {Mini Euso Experiment}},  {\em
  PoS} {\bf ICRC2023} (2023) 001.

\bibitem{Eser:2023Dw}
{\bf JEM-EUSO} Collaboration, J.~Eser, A.~V. Olinto, and L.~Wiencke, {\it
  {Overview and First Results of EUSO-SPB2}},  {\em PoS} {\bf ICRC2023} (2023)
  397.

\bibitem{Filippatos:2023P4}
{\bf JEM-EUSO} Collaboration, G.~Filippatos, {\it {EUSO-SPB2 Fluorescence
  Telescope in-flight performance and preliminary results}},  {\em PoS} {\bf
  ICRC2023} (2023) 251.

\bibitem{He:2023Hq}
{\bf LHAASO} Collaboration, H.~He, M.~Chen, C.~Hou, S.~Zhang, and X.~Zuo, {\it
  {Performances of the LHAASO detectors}},  {\em PoS} {\bf ICRC2023} (2023)
  416.

\bibitem{Verpoest:2023Fm}
{\bf IceCube} Collaboration, S.~Verpoest, {\it {Multiplicity of TeV muons in
  extensive air showers detected with IceTop and IceCube}},  {\em PoS} {\bf
  ICRC2023} (2023) 207.

\bibitem{Shefali:2023GK}
{\bf IceCube} Collaboration, S.~Shefali and F.~Schroeder, {\it {Status and
  plans for the instrumentation of the IceCube Surface Array Enhancement}},
  {\em PoS} {\bf ICRC2023} (2023) 342.

\bibitem{Noferini:2023OP}
{\bf EEE} Collaboration, F.~Noferini, {\it {Recent results from the
  PolarquEEEst measurement campaign at large geographical latitudes}},  {\em
  PoS} {\bf ICRC2023} (2023) 232.

\bibitem{Kling2023}
F.~Kling, {\it {The Forward Physics Facility and their implications for
  astroparticle physics}},  {\em PoS} {\bf ICRC2023} (2023) 023.

\bibitem{Morishima2023}
K.~Morishima, {\it {Cosmic ray imaging with nuclear emulsion plates for
  investigation of archaeological ruins}},  {\em PoS} {\bf ICRC2023} (2023)
  006.

\bibitem{Chai:20239v}
{\bf MAGIC} Collaboration, Y.~Chai, {\it {The cosmic-ray electron energy
  spectrum measured with the MAGIC telescopes}},  {\em PoS} {\bf ICRC2023}
  (2023) 323.

\bibitem{deNaurois:2023Hx}
{\bf H.E.S.S.} Collaboration, M.~de~Naurois, {\it {The Very-High-Energy
  electron spectrum observed with H.E.S.S.}},  {\em PoS} {\bf ICRC2023} (2023)
  261.

\bibitem{Xiong:2023Mz}
{\bf LHAASO} Collaboration, Z.~Xiong, S.~Wu, and H.~He, {\it {Measurement of
  cosmic-ray electrons with LHAASO KM2A-WCDA synergy}},  {\em PoS} {\bf
  ICRC2023} (2023) 315.

\bibitem{Morales-Soto:2023rX}
{\bf HAWC} Collaboration, J.~A. Morales-Soto, J.~C. Arteaga-Velázquez, and
  H.~Collaboration, {\it {HAWC measurements on the total energy spectrum of
  cosmic rays}},  {\em PoS} {\bf ICRC2023} (2023) 364.

\bibitem{ArteagaVelazquez:20238c}
{\bf HAWC} Collaboration, J.~C. Arteaga~Velazquez and H.~Collaboration, {\it
  {Analysis of the composition of TeV cosmic rays with HAWC}},  {\em PoS} {\bf
  ICRC2023} (2023) 299.

\bibitem{Vaidyanathan:2023wx}
{\bf TAIGA} Collaboration, A.~Vaidyanathan, {\it {The TAIGA-1 - A hybrid
  complex for gamma-ray astronomy, cosmic ray physics and astroparticle
  physics}},  {\em PoS} {\bf ICRC2023} (2023) 269.

\bibitem{Mohanty:2023fZ}
{\bf GRAPES-3} Collaboration, P.~Mohanty, {\it {Recent results from the
  GRAPES-3 experiment}},  {\em PoS} {\bf ICRC2023} (2023) 535.

\bibitem{Varsi:2023Sc}
{\bf GRAPES-3} Collaboration, F.~Varsi, {\it {Cosmic ray proton energy spectrum
  below the Knee observed by the GRAPES-3 experiment}},  {\em PoS} {\bf
  ICRC2023} (2023) 520.

\bibitem{Takita:20233f}
{\bf Tibet AS$\gamma$} Collaboration, M.~Takita, {\it {Highlights from the
  Tibet AS$\gamma$ experiment}},  {\em PoS} {\bf ICRC2023} (2023) 213.

\bibitem{Katayose:2023R1}
{\bf Tibet AS$\gamma$} Collaboration, Y.~Katayose, {\it {Measurement of the
  primary cosmic-ray proton spectrum between 40 TeV and a few hundred TeV with
  the Tibet hybrid experiment (Tibet=III + MD)}},  {\em PoS} {\bf ICRC2023}
  (2023) 301.

\bibitem{Zhang:2023fo}
{\bf LHAASO} Collaboration, H.~Zhang, H.~He, and C.~Feng, {\it {The
  all-particle spectrum and mean logarithmic mass of cosmic rays in the knee
  region measured with LHAASO-KM2A}},  {\em PoS} {\bf ICRC2023} (2023) 461.

\bibitem{Tian:2023cD}
{\bf LHAASO} Collaboration, X.~Tian, {\it {Probing cosmic ray composition with
  inclined air showers of LHAASO-KM2A}},  {\em PoS} {\bf ICRC2023} (2023) 297.

\bibitem{Rawlins:2023pw}
{\bf IceCube} Collaboration, K.~Rawlins, {\it {Accounting for changing snow
  over 10 years of IceTop, and its impact on the all-particle cosmic ray
  spectrum}},  {\em PoS} {\bf ICRC2023} (2023) 377.

\bibitem{Kang:20231c}
{\bf KASCADE-Grande} Collaboration, D.~Kang, {\it {Latest Analysis Results from
  the KASCADE-Grande Data}},  {\em PoS} {\bf ICRC2023} (2023) 307.

\bibitem{AbuZayyad:2023gk}
{\bf Telescope Array} Collaboration, T.~AbuZayyad, {\it {Cosmic ray energy
  spectrum and mass composition with the TALE fluorescence detector}},  {\em
  PoS} {\bf ICRC2023} (2023) 379.

\bibitem{Oshima:2023qJ}
{\bf Telescope Array} Collaboration, H.~Oshima, {\it {Measurement of cosmic-ray
  energy spectrum with the TALE detector in hybrid mode}},  {\em PoS} {\bf
  ICRC2023} (2023) 271.

\bibitem{Komae:2023R9}
{\bf Telescope Array} Collaboration, I.~Komae, {\it {Measurement of the cosmic
  ray energy spectrum in the 2nd knee region with the TALE-SD array}},  {\em
  PoS} {\bf ICRC2023} (2023) 405.

\bibitem{BrichettoOrquera:202340}
{\bf Pierre Auger} Collaboration, G.~Brichetto~Orquera, {\it {The second knee
  in the cosmic ray spectrum observed with the surface detector of the Pierre
  Auger Observatory}},  {\em PoS} {\bf ICRC2023} (2023) 398.

\bibitem{Ogio:2023S0}
{\bf Telescope Array} Collaboration, S.~Ogio, {\it {A study of the systematic
  effects on the energy scale for the measurement of UHECR spectrum by the TA
  SD array}},  {\em PoS} {\bf ICRC2023} (2023) 400.

\bibitem{Tsunesada:2023c0}
{\bf Pierre Auger and Telescope Array} Collaboration, Y.~Tsunesada, {\it
  {Measurement of UHECR energy spectrum with the Pierre Auger Observatory and
  the Telescope Array}},  {\em PoS} {\bf ICRC2023} (2023) 406.

\bibitem{Deligny:2023RB}
O.~Deligny, I.~Lhenry-Yvon, Q.~Luce, M.~Roth, D.~Schmidt, and A.~Watson, {\it
  {Energy dependence of the optimal distance used to determine the size of air
  showers: implications for the energy spectrum of ultra-high-energy cosmic
  rays}},  {\em PoS} {\bf ICRC2023} (2023) 533.

\bibitem{Vicha:2023px}
J.~Vicha, V.~Novotný, and J.~Ebr, {\it {Data-driven Estimation of Invisible
  Energy below EeV}},  {\em PoS} {\bf ICRC2023} (2023) 497.

\bibitem{Mayotte:2023WC}
{\bf Pierre Auger and Telescope Array} Collaboration, S.~Mayotte, {\it
  {Auger@TA: An Auger-like surface detector micro-array embedded within the
  Telescope Array Project}},  {\em PoS} {\bf ICRC2023} (2023) 368.

\bibitem{Mayotte:2023Nc}
{\bf Pierre Auger} Collaboration, E.~W. Mayotte, {\it {Measurement of the mass
  composition of ultra-high-energy cosmic rays at the Pierre Auger
  Observatory}},  {\em PoS} {\bf ICRC2023} (2023) 365.

\bibitem{ArteagaVelazquez:20231i}
{\bf KASCADE-Grande} Collaboration, J.~C. Arteaga~Velazquez, {\it {Energy
  dependence of the number of muons for hadronic air showers with
  KASCADE-Grande}},  {\em PoS} {\bf ICRC2023} (2023) 376.

\bibitem{Fujita:2023r8}
{\bf Telescope Array} Collaboration, K.~Fujita, {\it {Cosmic ray mass
  composition measurement with the TALE hybrid detector}},  {\em PoS} {\bf
  ICRC2023} (2023) 401.

\bibitem{Buitink:2023qO}
S.~Buitink et~al., {\it {High-resolution air shower observations with the
  Square Kilometer Array}},  {\em PoS} {\bf ICRC2023} (2023) 503.

\bibitem{Glombitza:2023}
{\bf Pierre Auger} Collaboration, J.~Glombitza, {\it {Mass Composition from 3
  EeV to 100 EeV using the Depth of the Maximum of Air-Shower Profiles
  Estimated with Deep Learning using Surface Detector Data of the Pierre Auger
  Observatory}},  {\em PoS} {\bf ICRC2023} (2023) 278.

\bibitem{Tkachenko:2023}
{\bf Pierre Auger} Collaboration, O.~Tkachenko, {\it {Studies of the mass
  composition of cosmic rays and proton-proton interaction cross-sections at
  ultra-high energies with the Pierre Auger Observatory}},  {\em PoS} {\bf
  ICRC2023} (2023) 438.

\bibitem{Stadelmaier:2023wr}
M.~Stadelmaier, V.~Novotn\'{y}, and J.~Vicha, {\it {The ability of the Greisen
  function to describe air shower profiles}},  {\em PoS} {\bf ICRC2023} (2023)
  340.

\bibitem{Bellido:2023Na}
{\bf Pierre Auger} Collaboration, J.~Bellido, {\it {The Fitting Procedure for
  Longitudinal Shower Profiles Observed with the Fluorescence Detector of the
  Pierre Auger Observatory}},  {\em PoS} {\bf ICRC2023} (2023) 211.

\bibitem{Harvey:2023w3}
{\bf Pierre Auger} Collaboration, V.~M. Harvey, {\it {A new cross-check and
  review of aerosol attenuation measurements at the Pierre Auger Observatory}},
   {\em PoS} {\bf ICRC2023} (2023) 300.

\bibitem{Yushkov:2023}
{\bf Pierre Auger and Telescope Array} Collaboration, A.~Yushkov, {\it {Depth
  of maximum of air-shower profiles: testing the compatibility of the
  measurements at the Pierre Auger Observatory and the Telescope Array}},  {\em
  PoS} {\bf ICRC2023} (2023) 249.

\bibitem{Gonzalez:2023}
{\bf Pierre Auger} Collaboration, N.~Gonz\'{a}lez, {\it {Search for primary
  photons at tens of PeV with the Pierre Auger Observatory}},  {\em PoS} {\bf
  ICRC2023} (2023) 238.

\bibitem{Kharuk:2023T1}
{\bf Telescope Array} Collaboration, I.~Kharuk, G.~Rubtsov, and M.~Kuznetsov,
  {\it {Search for EeV photon-induced events at the Telescope Array}},  {\em
  PoS} {\bf ICRC2023} (2023) 324.

\bibitem{Takahashi:2023Hg}
{\bf Telescope Array} Collaboration, K.~Takahashi, {\it {TA SD analysis for
  inclined air showers}},  {\em PoS} {\bf ICRC2023} (2023) 306.

\bibitem{Franco:2023}
{\bf Pierre Auger} Collaboration, D.~de~Oliveira~Franco, {\it {Search for
  evidence of neutron fluxes using Pierre Auger Observatory data}},  {\em PoS}
  {\bf ICRC2023} (2023) 246.

\bibitem{Liu:2023/H}
{\bf LHAASO} Collaboration, W.~Liu, {\it {Measurement of cosmic-ray
  anisotropies using LHAASO-WCDA}},  {\em PoS} {\bf ICRC2023} (2023) 186.

\bibitem{Gao:2023E3}
{\bf LHAASO} Collaboration, W.~Gao, {\it {The large-scale anisotropy of cosmic
  rays based on LHAASO-KM2A}},  {\em PoS} {\bf ICRC2023} (2023) 478.

\bibitem{McNally:2023Ma}
{\bf IceCube} Collaboration, F.~McNally, {\it {Cosmic Ray Anisotropy with
  Eleven Years of IceCube Data}},  {\em PoS} {\bf ICRC2023} (2023) 360.

\bibitem{Chakraborty:2023+X}
{\bf GRAPES-3} Collaboration, M.~Chakraborty, {\it {Small-scale anisotropy in
  the cosmic ray flux observed by GRAPES-3 at TeV energies}},  {\em PoS} {\bf
  ICRC2023} (2023) 513.

\bibitem{Yuan:2023}
Q.~Yuan, {\it {Understanding the spectra and anisotropies of Galactic cosmic
  rays}},  {\em PoS} {\bf ICRC2023} (2023) 202.

\bibitem{Golup:2023}
{\bf Pierre Auger} Collaboration, G.~Golup, {\it {An update on the arrival
  direction studies made with data from the Pierre Auger Observatory}},  {\em
  PoS} {\bf ICRC2023} (2023) 252.

\bibitem{Caccianiga:2023}
{\bf Pierre Auger and Telescope Array} Collaboration, L.~Caccianiga, {\it
  {Update on the searches for anisotropies in UHECR arrival directions with the
  Pierre Auger Observatory and the Telescope Array}},  {\em PoS} {\bf ICRC2023}
  (2023) 521.

\bibitem{Kuznetsov:2023}
{\bf Pierre Auger and Telescope Array} Collaboration, M.~Kuznetsov, {\it
  {Possible interpretations of the joint observations of UHECR arrival
  directions using data recorded at the Telescope Array and the Pierre Auger
  Observatory}},  {\em PoS} {\bf ICRC2023} (2023) 528.

\bibitem{PierreAuger:2022axr}
{\bf Pierre Auger} Collaboration, P.~Abreu et~al., {\it {Arrival Directions of
  Cosmic Rays above 32 EeV from Phase One of the Pierre Auger Observatory}},
  {\em Astrophys. J.} {\bf 935} (2022), no.~2 170,
  [\href{http://arxiv.org/abs/2206.13492}{{\tt 2206.13492}}].

\bibitem{Bister:2023}
{\bf Pierre Auger} Collaboration, T.~Bister, {\it {Constraining models for the
  origin of ultra-high-energy cosmic rays with spectrum, composition, and
  arrival direction data measured at the Pierre Auger Observatory}},  {\em PoS}
  {\bf ICRC2023} (2023) 258.

\bibitem{PierreAuger:2022atd}
{\bf Pierre Auger} Collaboration, A.~A. Halim et~al., {\it {Constraining the
  sources of ultra-high-energy cosmic rays across and above the ankle with the
  spectrum and composition data measured at the Pierre Auger Observatory}},
  {\em JCAP} {\bf 05} (2023) 024, [\href{http://arxiv.org/abs/2211.02857}{{\tt
  2211.02857}}].

\bibitem{Haung:2023}
{\bf Tibet AS$\gamma$} Collaboration, J.~Haung, {\it {Study of muons from high
  energy cosmic ray air showers measured with the Tibet hybrid experiment
  (YAC-II + Tibet-III + MD)}},  {\em PoS} {\bf ICRC2023} (2023) 508.

\bibitem{Cazon:2023ik}
L.~Cazon, H.~Dembinski, G.~Parente, F.~Riehn, and A.~A. Watson, {\it {The muon
  measurements of Haverah Park and their connection to the muon puzzle}},  {\em
  PoS} {\bf ICRC2023} (2023) 431.

\bibitem{Romanov:2023Z9}
{\bf KM3NeT} Collaboration, A.~Romanov and P.~Kalaczyński, {\it {Comparison of
  the atmospheric muon flux measured by the KM3NeT detectors with the CORSIKA
  simulation using the Global Spline Fit model}},  {\em PoS} {\bf ICRC2023}
  (2023) 338.

\bibitem{Stadelmaier:2023NZ}
{\bf Pierre Auger} Collaboration, M.~Stadelmaier, {\it {The number of muons
  measured in hybrid events detected by the Pierre Auger Observatory}},  {\em
  PoS} {\bf ICRC2023} (2023) 339.

\bibitem{ArteagaVelazquez:20236d}
{\bf EAS-MSU, IceCube, KASCADE-Grande, NEVOD-DECOR, Pierre Auger, SUGAR,
  Telescope Array, Yakutsk EAS Array and Haverah Park} Collaboration, J.~C.
  Arteaga~Velazquez, {\it {A report by the WHISP working group on the combined
  analysis of muon data at cosmic-ray energies above 1 PeV}},  {\em PoS} {\bf
  ICRC2023} (2023) 466.

\bibitem{Tiberio:2023of}
{\bf LHCf} Collaboration, A.~Tiberio, {\it {The LHCf experiment at the Large
  Hadron Collider: status and prospects}},  {\em PoS} {\bf ICRC2023} (2023)
  444.

\bibitem{Piparo:2023PB}
{\bf LHCf} Collaboration, G.~Piparo, {\it {Measurement of the very forward
  $\pi^0$ and $\eta$ meson productions in p-p collisions at $\sqrt{s}$=13 TeV
  with the LHCf detector}},  {\em PoS} {\bf ICRC2023} (2023) 447.

\bibitem{Soldin:2023hZ}
D.~Soldin, {\it {Astroparticle Physics with the Forward Physics Facility at the
  High-Luminosity LHC}},  {\em PoS} {\bf ICRC2023} (2023) 327.

\bibitem{Riehn:2023Wf}
F.~Riehn, R.~Engel, and A.~Fedynitch, {\it {Sibyll$^\bigstar$: ad-hoc
  modifications for an improved description of muon data in extensive air
  showers}},  {\em PoS} {\bf ICRC2023} (2023) 429.

\bibitem{Pierog:2023OB}
T.~Pierog and K.~Werner, {\it {EPOS LHC-R : up-to-date hadronic model for EAS
  simulations}},  {\em PoS} {\bf ICRC2023} (2023) 230.

\bibitem{Ebr:2023dR}
J.~Ebr, J.~Blazek, J.~Vicha, T.~Pierog, E.~Santos, P.~Travnicek, N.~Denner, and
  R.~Ulrich, {\it {Impact of modified characteristics of hadronic interactions
  on cosmic-ray observables for proton and nuclear primaries}},  {\em PoS} {\bf
  ICRC2023} (2023) 245.

\bibitem{Huege:2023}
{\bf CORSIKA 8} Collaboration, T.~Huege and M.~Reininghaus, {\it {The
  particle-shower simulation code CORSIKA 8}},  {\em PoS} {\bf ICRC2023} (2023)
  310.

\bibitem{Albrecht:20238R}
{\bf CORSIKA 8} Collaboration, J.~Albrecht, J.-M. Alameddine, and F.~Riehn,
  {\it {Validation of Electromagnetic Showers in CORSIKA 8}},  {\em PoS} {\bf
  ICRC2023} (2023) 393.

\bibitem{AlvesJr:2023Hz}
{\bf CORSIKA 8} Collaboration, A.~A. Alves~Jr, N.~Karastathis, and T.~Huege,
  {\it {Parallel processing of radio signals and detector arrays in CORSIKA
  8}},  {\em PoS} {\bf ICRC2023} (2023) 469.

\bibitem{Alameddine:2023Sp}
{\bf CORSIKA 8} Collaboration, N.~Karastathis, R.~Prechelt, J.~Ammerman-Yebra,
  M.~Reininghaus, and T.~Huege, {\it {Simulating radio emission from air
  showers with CORSIKA 8}},  {\em PoS} {\bf ICRC2023} (2023) 425.

\bibitem{Sako:2023Ox}
T.~Sako, {\it {Development of a general purpose air shower simulation tool
  COSMOS X}},  {\em PoS} {\bf ICRC2023} (2023) 294.

\bibitem{Baack:2023}
{\bf CORSIKA 8} Collaboration, D.~Baack, {\it {Comparison and efficiency of GPU
  accelerated optical light propagation in CORSIKA~8}},  {\em PoS} {\bf
  ICRC2023} (2023) 417.

\bibitem{Buckland:2023YC}
I.~Buckland and D.~Bergman, {\it {CHASM (CHerenkov Air Shower Model)}},  {\em
  PoS} {\bf ICRC2023} (2023) 325.

\bibitem{Krizmanic:2023O0}
J.~Krizmanic, J.~Mitchell, A.~Cummings, and F.~Garcia, {\it {Extensive Air
  Shower (EAS) Development in the Upper Atmosphere: a unique environment to
  measure the EAS properties}},  {\em PoS} {\bf ICRC2023} (2023) 524.

\bibitem{Tueros:2023rx}
M.~J. Tueros, S.~Cabana-Freire, and J.~Alvarez-Muniz, {\it {Radio-Emission from
  Atmosphere-Skimming Cosmic-Ray Showers in High-Altitude Balloon-Borne
  experiments}},  {\em PoS} {\bf ICRC2023} (2023) 349.

\bibitem{Chiche:2023oN}
S.~Chiche, {\it {New features in the radio-emission of very inclined
  air-showers}},  {\em PoS} {\bf ICRC2023} (2023) 394.

\bibitem{Schimassek:2023I6}
M.~L. Schimassek, R.~Engel, A.~Ferrari, M.~Roth, D.~Schmidt, and D.~Veberic,
  {\it {Simulations of neutrons in extensive air showers}},  {\em PoS} {\bf
  ICRC2023} (2023) 390.

\bibitem{Zhou:2023Vq}
{\bf LHAASO} Collaboration, X.~Zhou, C.~Yang, X.~Chen, and D.~Huang, {\it {Flux
  variations of cosmic ray air showers detected by LHAASO-KM2A during
  thunderstorms}},  {\em PoS} {\bf ICRC2023} (2023) 255.

\bibitem{Abbasi:2023Js}
{\bf Telescope Array and Lightning} Collaboration, R.~Abbasi, {\it {High-speed
  Video Camera Observations Associated with a Terrestrial Gamma-ray Flash at
  the Telescope Array Detector.}},  {\em PoS} {\bf ICRC2023} (2023) 250.

\bibitem{Colalillo:2023}
{\bf Pierre Auger} Collaboration, R.~Colalillo and J.~Dwyer, {\it {Study of
  downward Terrestrial Gamma-ray Flashes with the surface detector of the
  Pierre Auger Observatory}},  {\em PoS} {\bf ICRC2023} (2023) 439.

\bibitem{Mussa:2023}
{\bf Pierre Auger} Collaboration, R.~Mussa, {\it {Investigating multiple elves
  and halos above strong lightning with the fluorescence detectors of the
  Pierre Auger Observatory}},  {\em PoS} {\bf ICRC2023} (2023) 372.

\bibitem{Nakazawa:2023wo}
K.~Nakazawa et~al., {\it {Investigating the locations of electron acceleration
  in Hokuriku winter thunderclouds using on-ground gamma-ray and radio
  observations}},  {\em PoS} {\bf ICRC2023} (2023) 274.

\bibitem{Tsurumi:2023Zd}
M.~Tsurumi et~al., {\it {Lightning flash started near the electron acceleration
  region in the thundercloud}},  {\em PoS} {\bf ICRC2023} (2023) 254.

\bibitem{Wada:2023Y+}
Y.~Wada et~al., {\it {Lightning Mapping as a Probe of Electron Accelerator in
  Thunderclouds}},  {\em PoS} {\bf ICRC2023} (2023) 317.

\bibitem{SousaDiniz:2023O3}
G.~Sousa~Diniz et~al., {\it {Ambient conditions to reproduce gamma-ray glow
  energy spectra assuming cosmic ray as source}},  {\em PoS} {\bf ICRC2023}
  (2023) 209.

\bibitem{Tsuji:2023o}
N.~Tsuji et~al., {\it {Moon Moisture Targeting Observatory (MoMoTarO) for basic
  science application to neutron lifetime measurement}},  {\em PoS} {\bf
  ICRC2023} (2023) 296.

\bibitem{Romoli:2023C0}
{\bf JEM-EUSO} Collaboration, G.~Romoli, {\it {Study of multiple ring ELVES
  with the Mini-EUSO telescope on-board the International Space Station}},
  {\em PoS} {\bf ICRC2023} (2023) 223.

\bibitem{Nayak:2023J3}
{\bf GRAPES-3} Collaboration, P.~K. Nayak, {\it {Contemplating the observed
  relationship between the global electric circuit and GRAPES-3
  thunderstorm-induced muon events}},  {\em PoS} {\bf ICRC2023} (2023) 404.

\bibitem{Mulrey:2023v8}
K.~Mulrey et~al., {\it {Measuring cosmic rays with the LOFAR radio telescope}},
   {\em PoS} {\bf ICRC2023} (2023) 443.

\bibitem{Hariharan:2023Gw}
{\bf GRAPES-3} Collaboration, B.~Hariharan, {\it {Observation of the
  atmospheric wave created by Hunga Tonga-Hunga Ha’apai volcano eruption
  using GRAPES-3 detectors}},  {\em PoS} {\bf ICRC2023} (2023) 530.

\bibitem{Sandoval:20231J}
{\bf HAWC} Collaboration, A.~Sandoval, {\it {Observation of the Lamb wave
  created by the eruption of the Hunga volcano using cosmic rays detected by
  the HAWC observatory}},  {\em PoS} {\bf ICRC2023} (2023) 295.

\bibitem{Su:2023IJ}
S.-C. Su, Y.-C. Chen, J.~Nam, P.~Chen, and C.-Y. Kuo, {\it {Development of
  Affordable and Compact Muon Tomography Detector}},  {\em PoS} {\bf ICRC2023}
  (2023) 531.

\bibitem{Bhadra:2023Cu}
S.~Bhadra, {\it {Between the Cosmic Ray ‘knee’ and ‘ankle’ :
  Contribution from star clusters}},  {\em PoS} {\bf ICRC2023} (2023) 196.

\bibitem{Simeon:20233q}
P.~Simeon, N.~Globus, K.~Barrow, and R.~D. Blandford, {\it {Ultra-High-Energy
  Cosmic Rays from Accretion Shocks of Galaxy Clusters and Filaments}},  {\em
  PoS} {\bf ICRC2023} (2023) 369.

\bibitem{Peretti:2023aO}
E.~Peretti and M.~Ahlers, {\it {Particle acceleration and multi-messenger
  emission from ultra-fast outflows}},  {\em PoS} {\bf ICRC2023} (2023) 361.

\bibitem{Wang:2023ab}
J.~Wang, {\it {Shear acceleration in Active-Galactic-Nucleus jets}},  {\em PoS}
  {\bf ICRC2023} (2023) 194.

\bibitem{Araudo:2023ie}
A.~Araudo, {\it {Acceleration of UHECRs in AGN jets and backflows}},  {\em PoS}
  {\bf ICRC2023} (2023) 541.

\bibitem{Condorelli:2023z/}
A.~Condorelli, J.~Biteau, and O.~Deligny, {\it {Observational constraints on
  transient accelerators of ultra-high energy cosmic rays}},  {\em PoS} {\bf
  ICRC2023} (2023) 336.

\bibitem{Tomita:2023A0}
S.~Tomita and Y.~Ohira, {\it {Particle-in-cell Simulation of a Relativistic
  Shock Propagating in an Electron-Proton-Helium Plasma}},  {\em PoS} {\bf
  ICRC2023} (2023) 422.

\bibitem{Morikawa:2023te}
K.~Morikawa, {\it {Particle acceleration in a relativistic shock in
  inhomogeneous media}},  {\em PoS} {\bf ICRC2023} (2023) 378.

\bibitem{Orusa:20231g}
L.~Orusa and D.~Caprioli, {\it {Ion acceleration in 3D hybrid simulations of
  non-relativistic quasi-perpendicular shocks}},  {\em PoS} {\bf ICRC2023}
  (2023) 263.

\bibitem{Fulat:2023R+}
K.~Fulat, A.~Bohdan, G.~Torralba~Paz, M.~Tsirou, and M.~Pohl, {\it {PIC
  simulations of SNRs shocks with a turbulent upstream medium}},  {\em PoS}
  {\bf ICRC2023} (2023) 286.

\bibitem{Jikei:2023}
T.~Jikei, {\it {Simulation of Weibel instability in weakly magnetized
  astrophysical shocks}},  {\em PoS} {\bf ICRC2023} (2023) 485.

\bibitem{TorralbaPaz:2023y7}
G.~Torralba~Paz, A.~Bohdan, and J.~Niemiec, {\it {Prediction and Anomaly
  Detection of accelerated particles in PIC simulations using neural
  networks}},  {\em PoS} {\bf ICRC2023} (2023) 341.

\bibitem{Plotko:20237z}
P.~Plotko, A.~van Vliet, X.~Rodrigues, and W.~Winter, {\it {Differences between
  PAO and TA spectra: Systematics or indication of a local astrophysical
  source?}},  {\em PoS} {\bf ICRC2023} (2023) 229.

\bibitem{Ryu:2023}
D.~Ryu, {\it {Ultra-High Energy Cosmic Rays from Radio Galaxies}},  {\em PoS}
  {\bf ICRC2023} (2023) 210.

\bibitem{Taylor:2023RP}
A.~M. Taylor, J.~Matthews, and T.~Bell, {\it {UHECR Echoes from the Council of
  Giants}},  {\em PoS} {\bf ICRC2023} (2023) 215.

\bibitem{Watanabe:2023kX}
K.~Watanabe, A.~Fedynitch, F.~Capel, and H.~Sagawa, {\it {Overcoming Challenges
  in Finding Ultra-High-Energy Cosmic Ray Sources with a Bayesian Hierarchical
  Framework: Impact of the Galactic magnetic field and mass composition}},
  {\em PoS} {\bf ICRC2023} (2023) 479.

\bibitem{Bourriche:2023N/}
N.~Bourriche and F.~Capel, {\it {Cosmic cartography with UHECRs: Source
  constraints from individual events at the highest energies}},  {\em PoS} {\bf
  ICRC2023} (2023) 362.

\bibitem{Oikonomou:2023Ya}
F.~Oikonomou, D.~Ehlert, and M.~Unger, {\it {The Curious Case of the Maximum
  Rigidity Distribution of Ultra-high Energy Cosmic-Ray Accelerators}},  {\em
  PoS} {\bf ICRC2023} (2023) 240.

\bibitem{Farrar:2023vB}
G.~R. Farrar and T.~Bister, {\it {Anisotropies, large and small}},  {\em PoS}
  {\bf ICRC2023} (2023) 459.

\bibitem{Unger:2023Rn}
M.~Unger and G.~R. Farrar, {\it {New Models of the Magnetic Field of the
  Galaxy}},  {\em PoS} {\bf ICRC2023} (2023) 253.

\bibitem{Dolgikh:2023tq}
K.~A. Dolgikh, A.~Korochkin, G.~Rubtsov, D.~Semikoz, and I.~Tkachev, {\it
  {Caustic-like Structures in UHECR Flux after Propagation in Turbulent
  Intergalactic Magnetic Fields and the caused distortions of the image of a
  source}},  {\em PoS} {\bf ICRC2023} (2023) 452.

\bibitem{Higuchi:2023mL}
R.~Higuchi et~al., {\it {Influence of Galactic magnetic fields on UHECR energy
  spectrum and mass composition on the Earth}},  {\em PoS} {\bf ICRC2023}
  (2023) 460.

\bibitem{Globus:2023AK}
N.~Globus, A.~Fedynitch, and R.~D. Blandford, {\it {Extreme Energy Cosmic Rays
  "Treasure Maps": a new methodology to unveil the nature of cosmic
  accelerators}},  {\em PoS} {\bf ICRC2023} (2023) 440.

\bibitem{Olinto:2023ZQ}
{\bf POEMMA and JEM-EUSO} Collaboration, A.~V. Olinto, {\it {POEMMA (Probe Of
  Extreme Multi-Messenger Astrophysics) Roadmap Update}},  {\em PoS} {\bf
  ICRC2023} (2023) 1159.

\bibitem{Aloisio:2023K6}
{\bf NUSES} Collaboration, R.~Aloisio, {\it {The Terzina instrument on board
  the NUSES space mission}},  {\em PoS} {\bf ICRC2023} (2023) 391.

\bibitem{Bertaina:2023P6}
M.~E. Bertaina et~al., {\it {A new front end electronics for the detection of
  the optical Cherenkov signals by Extensive Air Showers directly observed from
  sub-orbital and orbital altitudes}},  {\em PoS} {\bf ICRC2023} (2023) 311.

\bibitem{Henrichs:2023XE}
{\bf RNO-G} Collaboration, A.~Nelles, {\it {Searching for cosmic-ray air
  showers with RNO-G}},  {\em PoS} {\bf ICRC2023} (2023) 259.

\bibitem{LeCoz:2023yz}
{\bf GRAND} Collaboration, S.~Le~Coz, {\it {Identification of air-shower radio
  pulses for the GRAND online trigger}},  {\em PoS} {\bf ICRC2023} (2023) 224.

\bibitem{Mitra:2023}
{\bf GRAND} Collaboration, P.~Mitra, {\it {Offline Signal Identification with
  GRANDProto300}},  {\em PoS} {\bf ICRC2023} (2023) 236.

\bibitem{Coleman:2023mP}
{\bf IceCube} Collaboration, A.~Coleman, {\it {The Surface Array of
  IceCube-Gen2}},  {\em PoS} {\bf ICRC2023} (2023) 205.

\bibitem{Kawata:2023mA}
{\bf ALPACA} Collaboration, K.~Kawata, {\it {First observational results of the
  ALPAQUITA air shower array in Bolivia}},  {\em PoS} {\bf ICRC2023} (2023)
  257.

\bibitem{Sakurai:20236q}
{\bf FAST} Collaboration, S.~Sakurai, {\it {Detecting ultra-high-energy cosmic
  rays with prototypes of the Fluorescence detector Array of Single-pixel
  Telescopes (FAST) in both hemispheres}},  {\em PoS} {\bf ICRC2023} (2023)
  302.

\bibitem{Tameda:2023vg}
{\bf CRAFFT} Collaboration, Y.~Tameda, {\it {Detector optimization and
  observation plan of the CRAFFT project for the next generation UHECR
  observation}},  {\em PoS} {\bf ICRC2023} (2023) 329.

\bibitem{AlvesBatista:20233O}
R.~Alves~Batista et~al., {\it {Science with the Global Cosmic-ray Observatory
  (GCOS)}},  {\em PoS} {\bf ICRC2023} (2023) 281.

\bibitem{Coleman:2022abf}
A.~Coleman et~al., {\it {Ultra high energy cosmic rays The intersection of the
  Cosmic and Energy Frontiers}},  {\em Astropart. Phys.} {\bf 149} (2023)
  102819, [\href{http://arxiv.org/abs/2205.05845}{{\tt 2205.05845}}].

\bibitem{Sarazin:2023Cz}
F.~Sarazin et~al., {\it {“Ultra-High-Energy Cosmic-Rays (UHECR): at the
  Intersection of the Cosmic and Energy Frontiers” -- Overview of the
  Snowmass UHECR white paper and roadmap}},  {\em PoS} {\bf ICRC2023} (2023)
  265.

\bibitem{Schroeder:2023uc}
F.~Schroeder et~al., {\it {Snowmass UHECR Whitepaper: Requirements on Future
  Instrumentation}},  {\em PoS} {\bf ICRC2023} (2023) 206.

\end{thebibliography}\endgroup
\bibliographystyle{JHEP}

\end{document}